\documentclass[11pt,a4paper]{article}
\usepackage{jheppub}
\usepackage{graphicx}
\usepackage{amssymb}
\usepackage{epstopdf}
\usepackage{subfigure}
\title{Identifying the colour of TeV-scale resonances}
\author[a]{S. Ask}
\author[a]{J.H. Collins}
\author[b]{J.R. Forshaw}
\author[b]{K. Joshi} 
\author[b,c]{A.D. Pilkington}
\affiliation[a]{Cavendish Laboratory, University of Cambridge, Cambridge CB3 0HE, UK.}
\affiliation[b]{School of Physics \& Astronomy, University of Manchester, Manchester M13 9PL, UK.}
\affiliation[c]{Institute of Particle Physics Phenomenology, University of Durham, Durham DH1 3LE, UK.}

\abstract{We explore how the colour of any new TeV-scale resonances
that decay into top quark pairs can be identified by studying the dependence of the observed cross-section on a central jet veto. 
To facilitate this study, colour octet resonance production was implemented 
in \pythia8{} and colour singlet resonance production is simulated
after minor modifications. We find 
that the colour of a 2~TeV resonance can be identified with 10~fb$^{-1}$ 
of data at a centre-of-mass energy of 14~TeV for a wide range of couplings, but only if the uncertainty in 
the theoretical prediction is dramatically reduced from its current level. 
}

\proceeding{Cavendish-HEP-11/17 \\ DCPT/11/84 \\ IPPP/11/42 \\ MAN/HEP/2011/13}

\newcommand{\eg}{{\it e.g.}}
\newcommand{\ie}{{\it i.e.}}
\def\pythia8{\ifmmode {\mbox{\textsc{Pythia8}}}\else
                 {\textsc{Pythia8}}\fi}%
\def\gkk{\ifmmode {\tilde{G}}\else
	  {$\tilde{G}$}\fi}%
\begin{document}
\maketitle

\section{Introduction}

The goal of this paper is to illustrate how one can gain insight into the 
colour of any new TeV-scale resonance that might be produced at the LHC. 
In the case where a resonance is observed in the invariant mass spectrum 
of dijets, the resonance mass will swiftly be measured as will its 
spin. However, it will also be important to identify the colour charge of the resonance. 
Generally speaking, differently coloured resonances will generate a different 
spectrum of accompanying radiation and one should expect to exploit this 
difference in order to establish the colour of the resonance. This matter 
has been explored in some recent papers \cite{Sung:2009iq,bib:gal10,bib:han10,Englert:2011cg} 
and our purpose is to present a detailed feasibility study that makes quantitative statements 
about the potential of such a measurement at the LHC. 

In this paper, we explore the associated radiation via 
the introduction of a jet veto. In particular, we compute the rate of 
resonance production subject to the constraint that there should be no 
jets (apart from those arising from the decay of the resonance) lying in the central region of 
rapidity with transverse momentum above some scale, $Q_0$. The variation of 
the cross section with $Q_0$ contains important information on the colour 
of the resonance. The method of jet vetoing to probe the colour structure of new physics should have 
other important applications, for example in \cite{Cox:2010ug} it was demonstrated 
that one can extract the effective couplings of the Higgs boson to weak vector 
bosons and to gluons by measuring the veto-scale dependence in ``Higgs plus dijet'' 
events with a veto on any third jet lying between the primary jets.
We therefore expect the method to have general utility. 

Jet vetoing has a number of advantages which make it well suited to
the task in hand. Experimentally, the systematic uncertainties are
expected to be small if one focusses on a ``gap fraction'' (i.e. the
ratio of the cross section with a veto to that without). This
expectation is born out by recent gap fraction measurements in dijet
events at the LHC \cite{bib:atl11}. Apart from that, any attempt to
identify the colour structure of new physics processes will be
hindered by background and by pile-up. One can minimize sensitivity to
background by using observables that are defined over the ensemble of
events and which can therefore be corrected for background provided it
can be subtracted statistically. The gap fraction is such
an observable since it is a measure of the $Q_0$ dependence of the
signal cross section. Observables that are measured on an
event-by-event basis would suffer from inevitable contamination from
background processes. In the high-luminosity phase of the LHC, there
will typically be more than 20 pile-up interactions overlaying the
signal and any method used to extract the colour flow needs to be
robust to this extra activity. The jet veto method will be robust against
pile-up if $Q_0$ is chosen to be large and/or the vetoed jets are
within the acceptance of the inner tracking detectors, as discussed in
Section~\ref{sec:veto}. 

It is possible to study the colour structure of events using other
methods. One such possibility is jet scaling
\cite{Englert:2011cg,Gerwick:2011tm}, which uses the
cross-section ratios, $\sigma(n+1)/\sigma(n)$, for specific jet
multiplicities, $n$. The information contained in these ratios is very similar to that contained in the gap fraction and we do not
discuss it any further here. Another possibility might be to exploit
the structure of jets to probe the colour flow, e.g. the `jet pull'
\cite{bib:gal10}. This approach should yield important information,
however it suffers from event-by-event contamination from background
and the effect of pile-up would also need to be studied. We shall focus
on the jet veto method in what follows, whilst acknowledging that
other methods may have a role to play should a new resonance be
observed at the LHC.

Resonant production of a new heavy gauge boson is the primary signal 
of new physics at hadron colliders in several extensions of the Standard Model 
(SM). One popular scenario is within the Randall-Sundrum (RS) model 
\cite{bib:ran99,bib:lil07,bib:aga08} where all the SM fields have access to 
the extra dimensional space. In this scenario, the lightest Kaluza-Klein (KK) 
excitations of the gauge bosons primarily decay into top quarks. There are 
several alternative scenarios within this RS framework which have attracted 
interest as candidates to resolve a number of issues 
with the SM, \eg\ the gauge hierarchy problem \cite{bib:ran99}, 
the fermion mass hierarchy \cite{bib:gro00,bib:ghe00,bib:hub00}, 
gauge coupling unification \cite{bib:ran01,bib:aga03,bib:car03,bib:aga05}, 
providing a dark matter candidate \cite{bib:aga04,bib:aga052}. 
In addition, recent results on the forward-backward asymmetry in top 
pair events from the CDF and D\O\ experiments indicate a potential deviation from 
the SM expectation \cite{Aaltonen:2011kc,Aaltonen:2008hc,d0:2011rq,d0:2007qb}, which 
could be caused by a colour octet resonance with a mass around 2 TeV \cite{Bai:2011ed}. 
For these reasons, we have chosen to investigate how the veto-scale dependence can 
discriminate between a colour octet and a colour singlet resonance,
both with a mass of 2~TeV and spin-1.\footnote{No other colour is possible if the resonance
  is to couple to quark--anti-quark pairs.} 
  
The outline of the paper is as follows: In Section 2, we discuss a new
implementation of heavy gluon resonances in the \pythia8 Monte Carlo
event generator. In Section 3, we use this implementation to study the
feasibility of identifying the colour of such a resonance at the
LHC. We start from the assumption that a new resonance has been
observed at the LHC in $pp$ collisions at $\sqrt{s} = 14$~TeV. We then
show how to quantify the probability of
sucessfully extracting the colour of the resonance. Finally, we
discuss the limiting theoretical and experimental uncertainties
associated with the measurement.

\section{Implementation of a gluon resonance in \pythia8}\label{sec:pytmodel}

We have implemented the heavy gluon resonance process within the Monte 
Carlo generator \pythia8\ \cite{Sjostrand:2007gs,bib:sjo06}. The implementation 
assumes that the heavy gluon can only be produced by quarks, i.e. 
$q\bar{q} \rightarrow \gkk \rightarrow q \bar{q}$. At tree-level, the suppression of any
coupling to gluons is a consequence of the Landau-Yang theorem (for
massless gluons and a heavy spin-1 resonance)  
\cite{Landau,Yang:1950rg} and it has also been shown that the suppression
persists at one-loop in the Bulk RS model \cite{bib:all10}. We include 
interference with SM QCD $2 \to 2$ production and allow for general couplings 
to quarks, in particular different couplings to left and right-handed
states are permitted. 
Such chiral couplings have recently been discussed in connection with the forward-backward 
asymmetry ($A_{FB}$) measurements at the Tevatron \cite{bib:djo09,bib:fer09}. 
A forward-backward asymmetry caused by the heavy gluon at leading-order 
requires an axial-vector coupling to both the initial and final state quarks. 
For this reason, such an asymmetry is not expected in the common RS models 
\cite{bib:bau10}, where the light quark couplings are vector-like. However, such 
studies might be of interest in alternative models.

Although the implementation is generic, we shall continue to make reference to the 
most common RS models, where the KK gluon coupling to top quarks is significantly 
stronger than to the other flavours and the total width is therefore dominated by 
the width from decays into tops. The fact that the final state flavour is normally 
different from the initial state flavour has the consequence that the light-quark 
couplings approximately determine the production cross section whereas the top coupling 
fixes the total width.  The $b$-quark coupling is often sufficiently small that 
it only weakly effects the width and cross section, we therefore restrict the parameter 
space by fixing $g^b_v$ to either be zero or the reference RS model value \cite{bib:lil07} 
discussed below. 
The heavy gluon uses the particle id code 5100021 and its corresponding decay 
channels can be specified through the standard \pythia8 particle data scheme. 

Specifically, the total production cross section is given by
\begin{eqnarray}
\sigma(\hat{s}) &=& \frac{8 \pi  \alpha^2_S  }{27} \hat{s}\left[\sigma_{SM} + \sigma_{KK} + \sigma_{int} \right] ~,\\
\sigma_{SM} &=& \frac{\beta_j A_j}{\hat{s}^2} ~,\\
\sigma_{KK} &=& \frac{((g^i_v)^2+(g^i_a)^2) ((g^j_v)^2 \beta_j A_j + (g^j_a)^2 \beta^3_j)}
{(\hat{s}-m^2_{\gkk})^2 + \left ( \hat{s} \frac{\Gamma_{tot}}{m_{\gkk}} \right )^2 }~, \\
\sigma_{int} &=& \frac{2}{\hat{s} } \cdot \frac{g^i_vg^j_v (\hat{s}-m^2_{\gkk}) \beta_j A_j}
{(\hat{s}-m^2_{\gkk})^2 + \left ( \hat{s} \frac{\Gamma_{tot}}{m_{\gkk}} \right )^2 }~,
\label{eq:int}
\end{eqnarray}
where the subscripts indicate the contributions from the 
SM and KK gluon amplitudes and their interference. In the above formulae $g^{i/j}_v$ 
and $g^{i/j}_a$ represent the vector and axial-vector couplings to the initial and 
final state quarks, with flavours $i$ and $j$; the couplings are all 
relative to the strong coupling in accordance with \cite{bib:dav01}. The partonic 
centre-of-mass energy is denoted $\hat{s}$ and the quantities $\beta_j$ and $A_j$ 
are defined via
\begin{eqnarray}
\beta_j = \sqrt{1 - 4\frac{m_j^2}{\hat{s}}} ~,\\
A_j = 1 + 2\frac{m_j^2}{\hat{s}}.
\end{eqnarray}
The masses $m_{\gkk}$ and $m_j$ correspond to the KK gluon (which we
generally refer to as the ``heavy gluon'') and final state quark 
masses and the total width is obtained by summing the partial widths:
\begin{eqnarray}
\Gamma(\gkk \rightarrow q_j \bar{q}_j) = \frac{\alpha_S  \beta  \sqrt{\hat{s}}}{6} 
\left [ (g^j_v)^2 A_j + (g^j_a)^2\beta^2 \right ]~.
\end{eqnarray}

In addition to specifying the KK gluon mass, the program offers the possibility 
to assign separate values of the KK gluon coupling to light, bottom
and top quarks. The most common RS models often predict different couplings 
to different helicity states and therefore the couplings can also be assigned
separately for left-handed and right-handed quarks. Finally, the hard 
process is integrated within the \pythia8\ framework of parton showers, underlying 
event activity, hadronisation and unstable particle decays. The
implementation has been validated 
against results in the literature and it has been used to investigate characteristic 
features of a heavy colour octet: some of those results are presented 
and discussed in the appendix.

\section{Extracting the colour of a heavy resonance}

In this section, we investigate the feasibility of distinguishing a 2~TeV colour 
octet resonance from an otherwise identical colour singlet resonance by vetoing 
on additional jet activity outside the di-top system.

\subsection{Simulation and event selection}

Events are generated using the  \pythia8{} implementation of heavy gluon resonance 
production, as discussed in Section \ref{sec:pytmodel}. The couplings used for the 
gluon resonance are $g^q_v = 0.2$, $g^b_v = 0$ and $g^t_v = 3.6$.  
The mass chosen for this study is 2~TeV. The resulting cross section $\times$ 
branching-ratio is 1.1~pb and the resonance width is $\Gamma / M = 0.2$. Purely 
vector couplings are assumed and interference with the SM contribution 
is ignored. 
Heavy photon production is simulated by changing the colour flow. Specifically, we 
replace the colour factors and  $\alpha_s \to e_q^2 \alpha_{em}$, where $e_q$ 
is the electric charge\footnote{In other words, we assume that the heavy colour 
singlet couples like a true photon.}. We also adjusted (i) the coupling of the heavy 
photon to light quarks to reproduce the production rate for the heavy gluon and (ii) 
the coupling of the heavy photon to top quarks in order to match the total width of 
the heavy gluon.  The {\tt CTEQ5L} parton distribution functions are used \cite{bib:lai00} with the 
default tune to the underlying event (UE) of version 8.130 ({\tt Tune1}) \cite{Sjostrand:2007gs}.
Details of the UE tune are not very important -- the sensitivity 
to non-perturbative physics was tested by repeating the analysis, but with hadronisation and 
multiple parton interactions switched off. We observe 
that the change in the gap fraction is less than 2\% and therefore
conclude that the soft physics modelling is not a crucial 
component in determining the feasibility of this measurement.

\begin{figure}[ht]%
\centering
\mbox{
\subfigure[]{\includegraphics[width=0.45\textwidth]{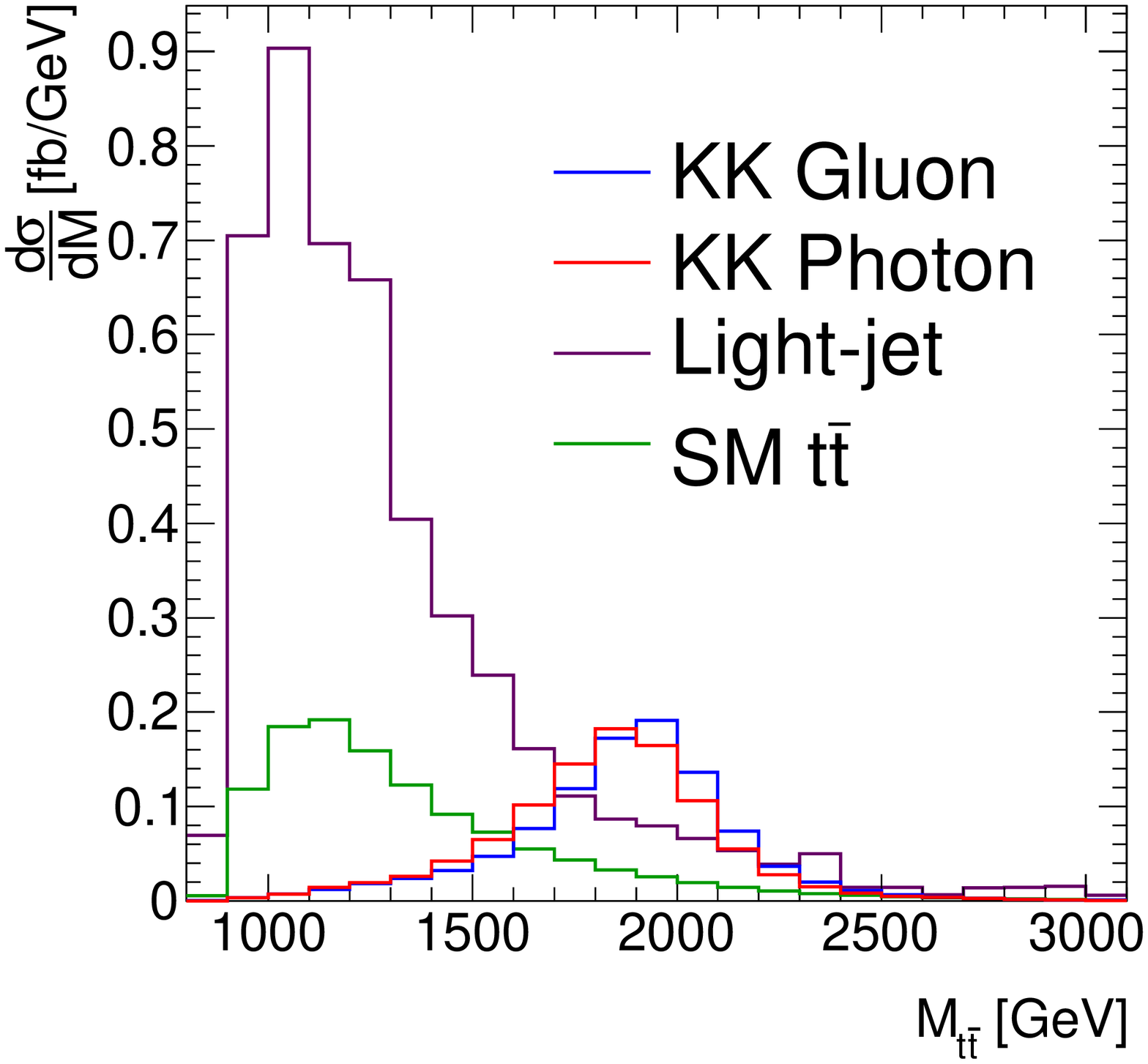}}\quad
\subfigure[]{\includegraphics[width=0.45\textwidth]{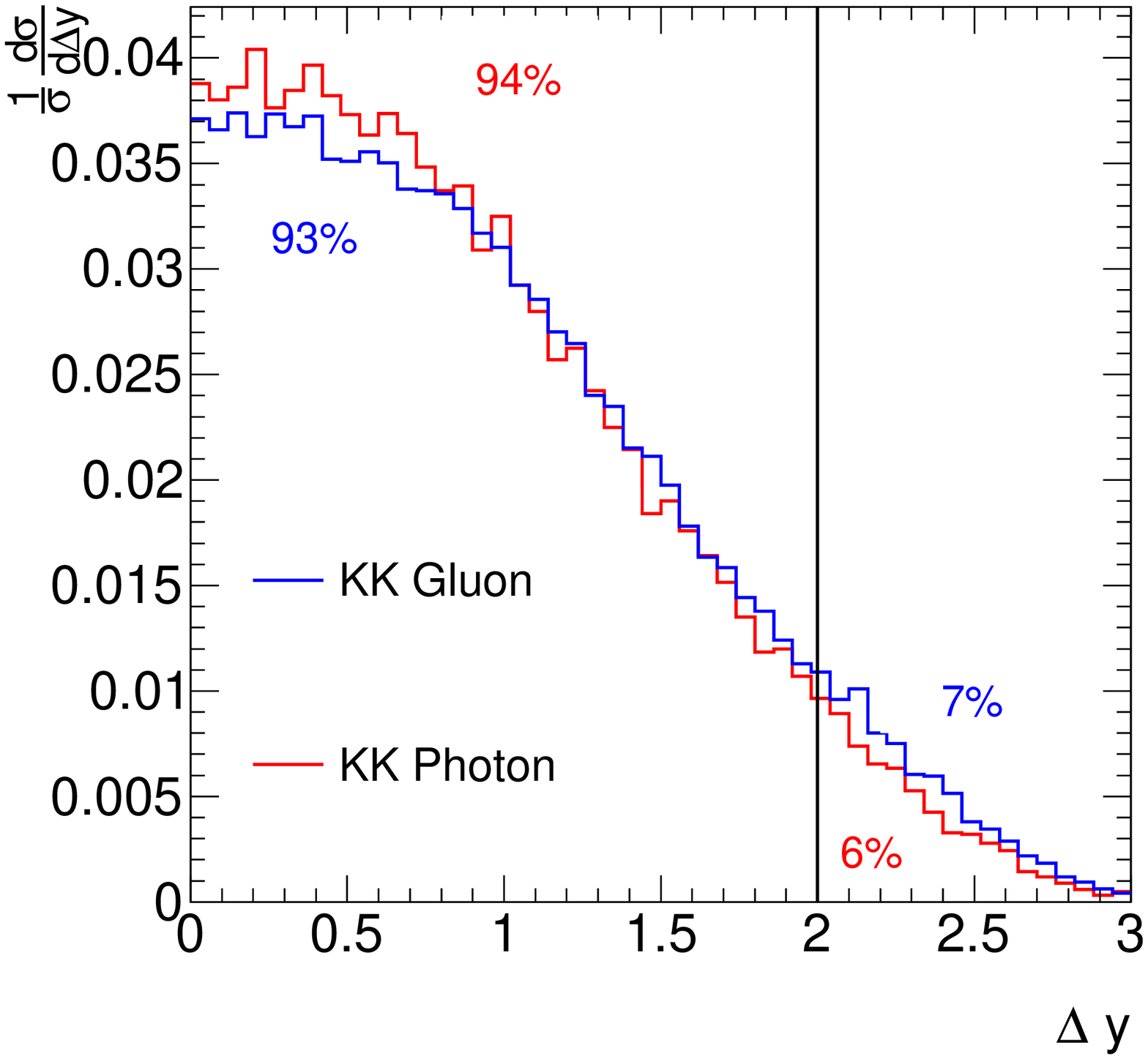}}
}
\caption{Signal and background distributions after tagging the two leading jets 
as top candidates. (a) The invariant mass of the top candidates. (b) The $\Delta y$ 
between the top candidates. 
\label{fig:sigbkd}}
\end{figure}

The top quarks will be highly boosted and form a 
dijet topology and so the primary background will come from QCD $2\rightarrow2$ 
scattering. We define $t\bar{t}$ and light-jet backgrounds separately, the latter 
being defined with no top-quarks in the final state. 
In both cases, the SM background events were simulated using \pythia8. 
Top-jet candidates are identified using the {\sc FastJet} library \cite{bib:cac08,bib:cac06,bib:cacweb} 
and the Johns Hopkins top-tagging algorithm \cite{bib:kap08} (the parameters 
of the top-tagging algorithm are listed in Table~\ref{tab:jetdefs}). The input to the algorithm 
are stable particles from the MC event record that have $|\eta| < 4.9$, i.e. within 
the fiducial acceptance of the LHC detectors (neutrinos are removed). The
events are then
required to contain two tagged top-jets, each of which has $p_{\rm
  T}>400$~GeV, after which the 
cross section for heavy gluon production is 46.8~fb and for heavy
photon production it is 31.4~fb.
This difference demonstrates an interesting and important difference in the tagging 
efficiency for top quarks originating from differently coloured
resonances. We have confirmed that this difference is not due to
non-perturbative physics, in particular we confirm that the
difference in tagging efficiency arises after the parton shower with
hadronization/underlying event playing a small role.
It follows that, should such a resonance be observed at the LHC, the measurement 
of the colour flow (using a method such as that presented here) will be crucial in determining
the production cross section.\footnote{The use of jet substructure
  (e.g. see the review in \cite{Abdesselam:2010pt})  should yield additional information regarding the colour of the
  resonance, assuming that the effects of background contamination and pile-up can be brought under control.}
For both resonances, the top tagging is sufficient to reduce the background 
to a manageable level in the region of the resonance. Figure \ref{fig:sigbkd}(a) 
shows the invariant mass distributions of the tagged top quarks for each type of 
signal and background process. 

In order to compare like-for-like resonances, we make a further
adjustment to the heavy photon cross section so that it is equal to
the heavy gluon cross section after top tagging, i.e. $\sigma_0=46.8$~fb. In
our final analysis,  all results will be presented for a range of
values of the production cross section (i.e. $\sigma_0$ is a
baseline value).

\begin{table}[t]
\centering
\begin{tabular}{c|c|c|c}
 $\sum E_{T}$ (GeV) & $\delta_p$ & $R_{\rm CA}$ & $\delta_r$ \\ \hline
$E_{T} \leq 1000$ & 0.13 & 0.9 & 0.19 \\
$1000 < E_{T} \leq 1600$ & 0.10 & 0.8 & 0.19 \\
$1600 < E_{T} \leq 2600$ & 0.05 & 0.6 & 0.19 \\
$E_{T} > 2600$ & 0.05 & 0.4 & 0.16 
\end{tabular}
\caption{Parameters of the top-tagging algorithm. $R_{\rm CA}$ is the distance 
parameter that is used in the Cambridge-Aachen jet finding algorithm. The distance 
parameter is defined on an event-by-event basis since it is dependent on the scalar 
summed $E_{T}$ of all particles in the event. The parameters $\delta_p$ and $\delta_r$ are defined in \cite{bib:kap08}.
\label{tab:jetdefs}}
\end{table}%

\subsection{Definition of the jet veto region and the gap-fraction}\label{sec:veto}

To identify the additional jet activity, we use the anti-$k_{\rm T}$ algorithm with 
$R=0.6$. The inputs to this second-stage jet finding are the full set of stable 
interacting particles, i.e. the same as to the top-tagging stage. We keep only 
those jets, $j$, that are sufficiently far in $\eta-\phi$ space from the previously 
tagged top jets, $t$, i.e.
\begin{equation}
\Delta R = \left(\Delta\eta_{j, t}^2 + \Delta\phi_{j, t}^2 \right)^{1/2} > R_{\rm CA}
\end{equation} 
where $R_{\rm CA}$ is defined on an event-by-event basis as discussed in Table 
\ref{tab:jetdefs}. A veto can then be applied to the remaining jets in order to 
elucidate the colour structure of the event.

It might be anticipated that the best separation between the colour singlet and the colour 
octet resonance would occur by vetoing jet activity between the two top-quark jets 
\cite{Sung:2009iq}. However, this requires the top-jets to be separated by a large 
enough rapidity interval. Figure \ref{fig:sigbkd}(b) shows the signal cross sections 
as a function of the rapidity separation and less than 10\% of signal events would be 
retained by a rapidity cut $\Delta y>2$. Fortunately, this efficiency loss can be 
avoided if the veto is applied to jets that lie in a {\it central
  rapidity interval}, i.e. 
$|y|<1.5$. This choice remains sensitive to the different colour
structures because colour octet resonances radiate preferentially in the forward/backward regions 
whereas colour singlet resonances radiate preferentially in the central region.

Using a central rapidity interval for vetoing has the added benefit that the jets are 
contained within the acceptance of the tracking systems of the LHC experiments. Even 
though the details of such criteria are beyond the scope of this study, this will be 
important in order to control effects from pile-up, which can be reduced by examining 
the in-jet track activity and determining the amount of this activity that comes from 
a single vertex. 

\begin{figure}[h!t]%
\centering
\mbox{
\subfigure[]{\includegraphics[width=0.33\textwidth]{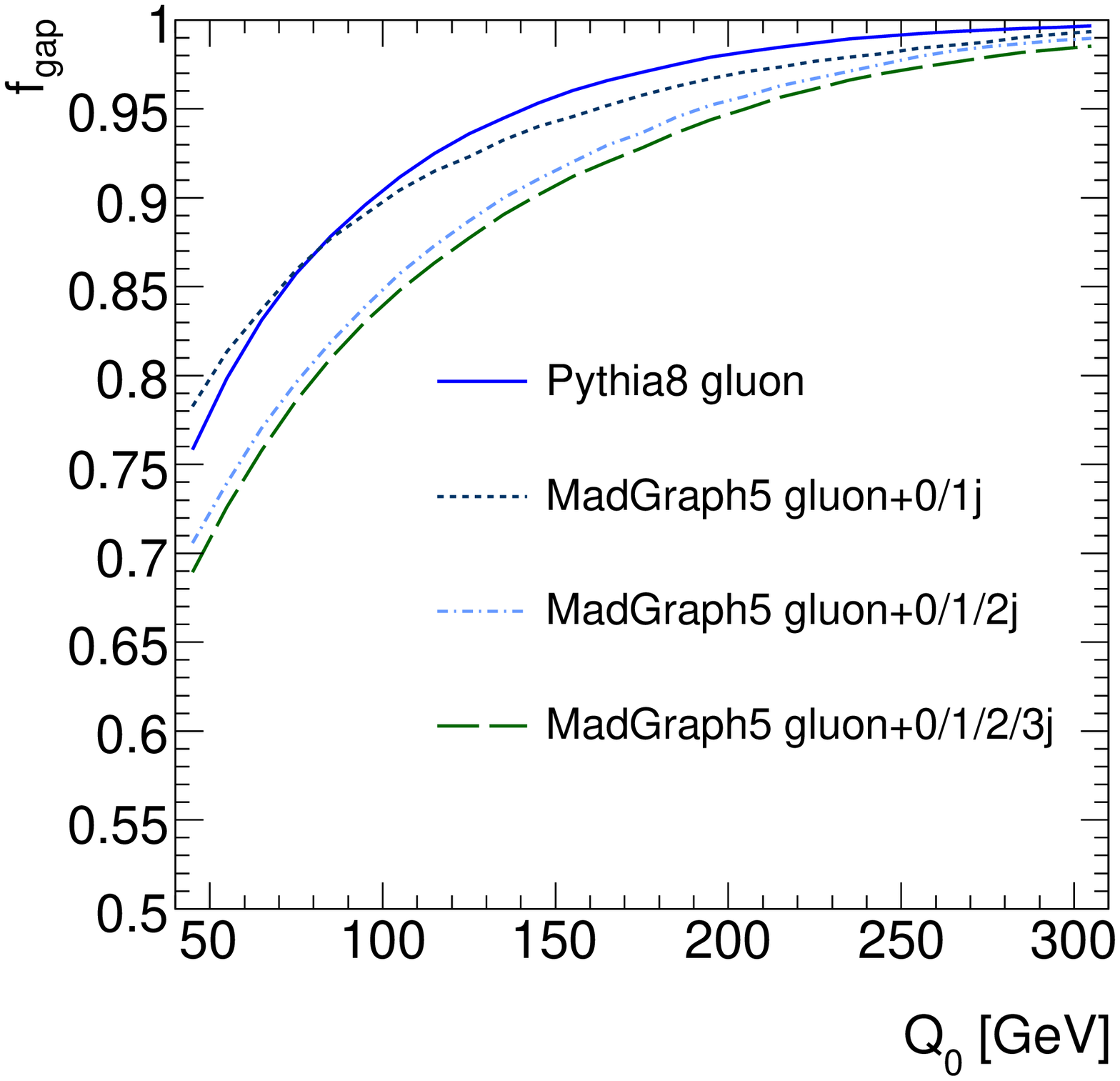} }%
\subfigure[]{\includegraphics[width=0.33\textwidth]{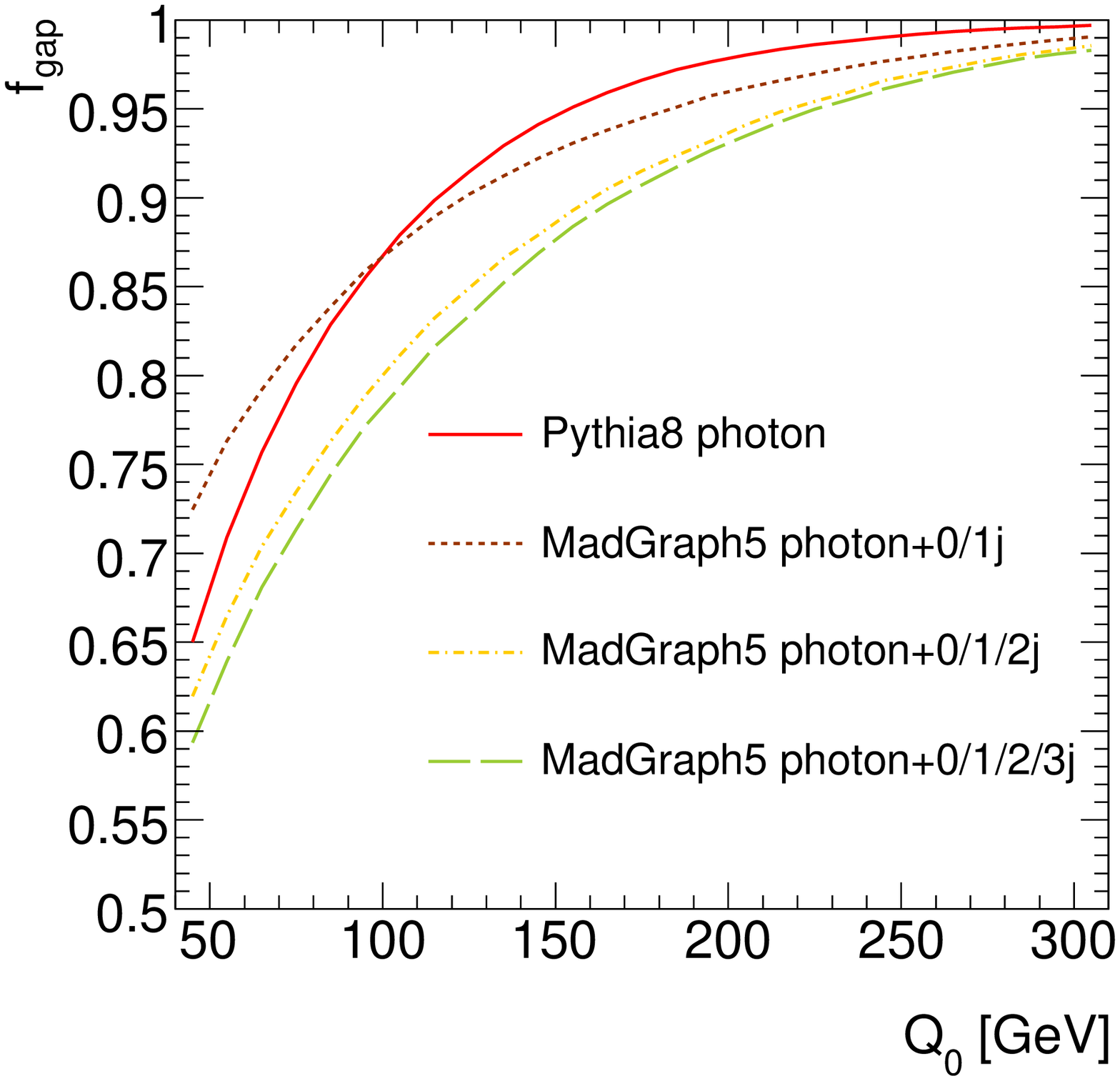} }%
\subfigure[]{\includegraphics[width=0.33\textwidth]{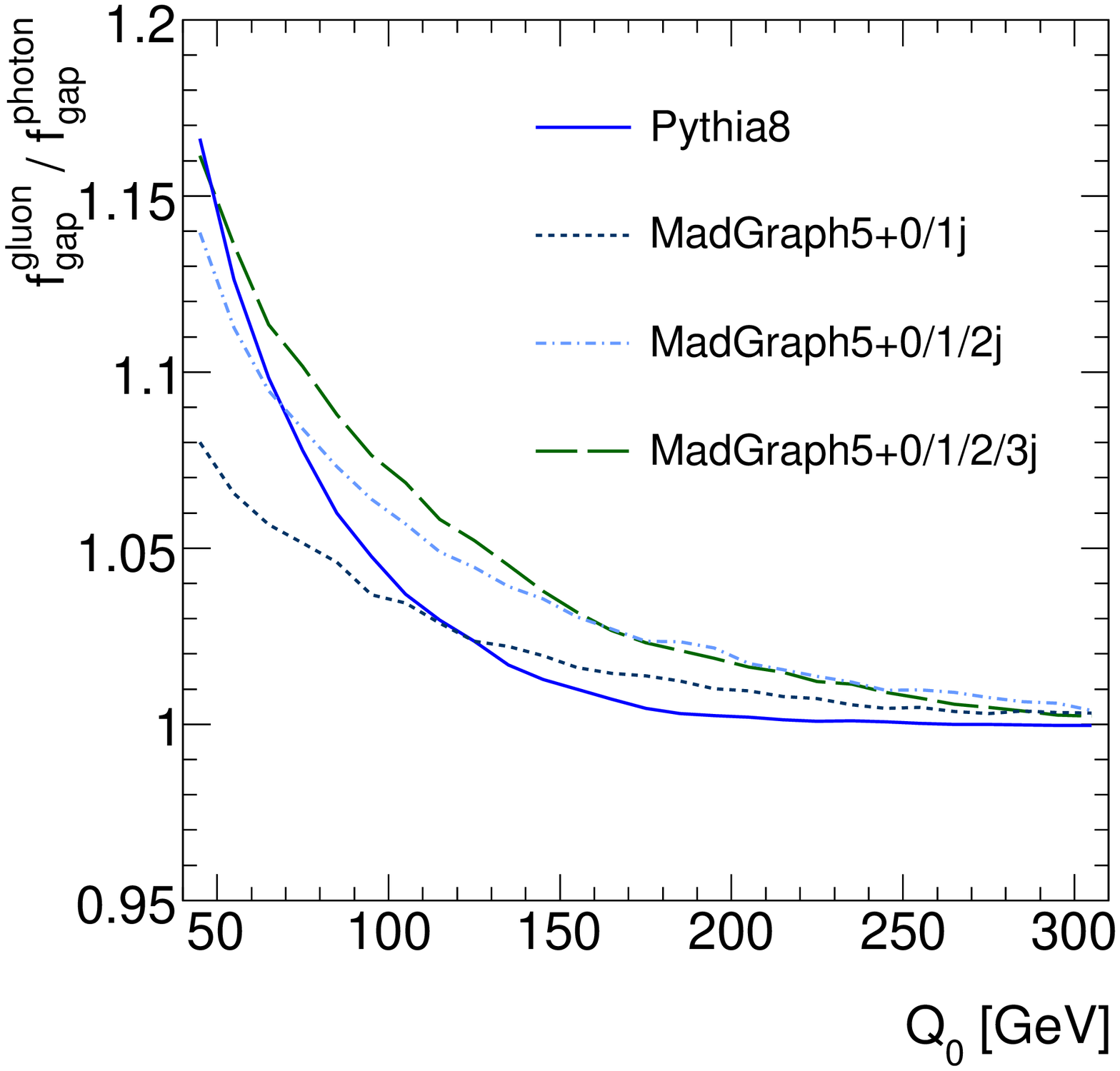} }%
}
\caption{The gap-fraction for heavy gluon (a) and heavy photon (b) resonances predicted by \pythia8 and MadGraph. The Madgraph curves correspond to 1,2 or 3 extra partons.
(c) The ratio of gluon and photon gap-fractions predicted by \pythia8 and MadGraph. 
\label{fig:gap-fracMG}}
\end{figure}

The observable of interest is the gap fraction, $f_{gap} (Q_0)$, defined as the fraction of the events that do not have a jet (in addition to the top-quark decay products) with $p_{\rm T}>Q_0$ and $|y|<1.5$, i.e.
\begin{equation} 
f_{gap} (Q_0) = \frac{\sigma(Q_0)}{\sigma(Q_{0}=300~\mbox{GeV})}.
\label{eq:fgap1}
\end{equation}
Figure \ref{fig:gap-fracMG}(a) shows the gap-fraction as a function of
$Q_0$ for a heavy gluon resonance predicted by \pythia8 and also by
the matrix element generator MadGraph/MadEvent 5
\cite{Alwall:2011uj}. The Madgraph predictions are determined for up
to three partons (in addition to the $t\bar{t}$) in the final
state. Figure \ref{fig:gap-fracMG}(b) shows the equivalent curves for
a heavy photon resonance. It is clear that the \pythia8 and MadGraph
predictions differ for each resonance by about 10\% at the lowest
values of $Q_0$. Although these theoretical predictions are
significantly different from each other (we discuss the theory uncertainty further in
Section \ref{sec:syst}), Figure \ref{fig:gap-fracMG}(c) shows that the
difference between the heavy gluon and heavy photon gap fractions 
remains sizeable in each case and this is all we need for the fidelity
of our analysis. We therefore use the results obtained using \pythia8 in
the following sections to determine the
sensitivity of the LHC experiments to the colour charge of new
resonances.

\subsection{Extracting the signal from background}

A pseudo-experiment approach is adopted in order to determine the expected
sensitivity to the colour of the resonance 
at a given luminosity, $L$. For each signal or background process, $i$, 
the expected number of events is determined by
$
\lambda_i = \sigma_i L, 
$
where $\sigma_{i}$ is the process cross section. The actual number of events, $n_i$, that contribute to a given pseudo-experiment 
is determined using a Poisson distribution with mean equal to $\lambda_i$. The pseudo-experiment is constructed by selecting these events at random from reduced MC samples, 
i.e. after top-tagging. 

The $t\bar{t}$ invariant mass distribution, $m_{t\bar{t}}$, is
constructed for several values of $Q_0$ (i.e. after vetoing 
events that contain additional jets with $p_{\rm T}^{} > Q_0$ in the rapidity interval 
$|y|<1.5$). From this we extract the size of the signal as a function of the veto scale for
$20$~GeV~$<Q_0<300$~GeV. The number of signal events at each value of $Q_0$ is determined 
by fitting the invariant mass distribution for both signal and background. 
The signal is parameterized by a skewed Breit-Wigner:
\begin{equation}
\frac{m_0^2 \Gamma^2 [ a + b(m_{t\bar{t}} - m_0)]}{(m_{t\bar{t}}^2 - m_0^2)^2 -
  m_0^2 \Gamma^2} 
\end{equation}
and $m_0$, $\Gamma$, $a$ and $b$ are allowed to vary in the fit. 
The background is assumed to have a shape determined by the $t\bar{t}$
sample and only its normalization is allowed 
to vary in the fit. Using just the $t\bar{t}$ background is reasonable since the
light-quark background has a very similar shape at large invariant
masses. Moreover, we have examined in detail what happens if we allow
the shape of the background fit to vary whilst keeping the
shape of the true background unchanged: Our final results
are very insensitive to even quite large changes in this shape and we are in any case confident that an experimental
analysis could reliably extract the signal cross section from background.
After fitting the invariant mass distribution, the number of signal events is given by $N_s = N_{T} - N_{b}$,
where $N_{T}$ is the total number of events in the pseudo-experiment 
that satisfy $1.5$~TeV~$\leq m_{t\bar{t}} <2.5$~TeV and $N_{b}$ is the corresponding number of 
background events in this range, as extracted from the fit. The size
and location of this mass window would, of course, be optimized in any
analysis.

\begin{figure}[h!t]%
\centering
\mbox{
\subfigure[]{\includegraphics[width=0.45\textwidth]{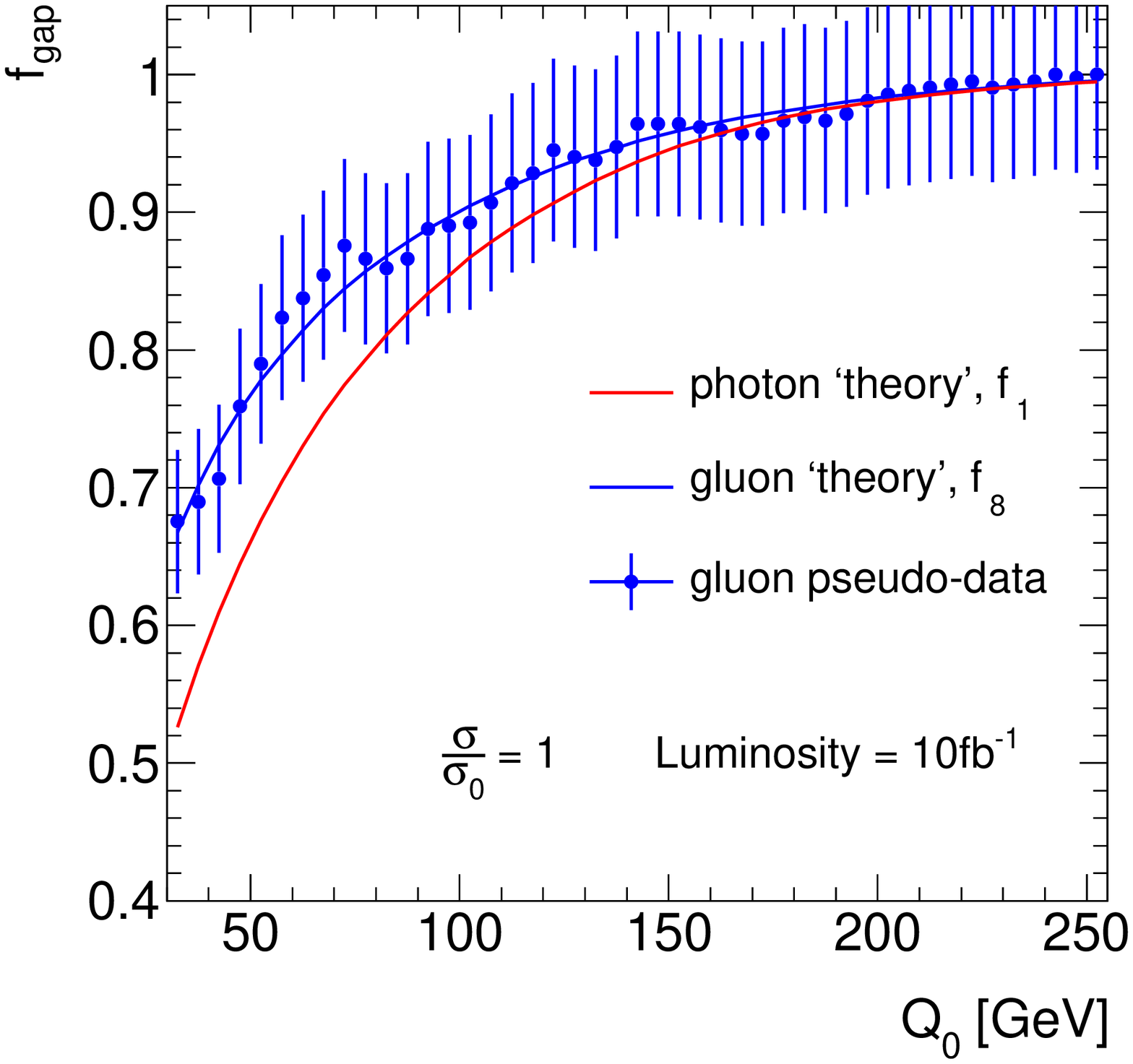} }%
\subfigure[]{\includegraphics[width=0.45\textwidth]{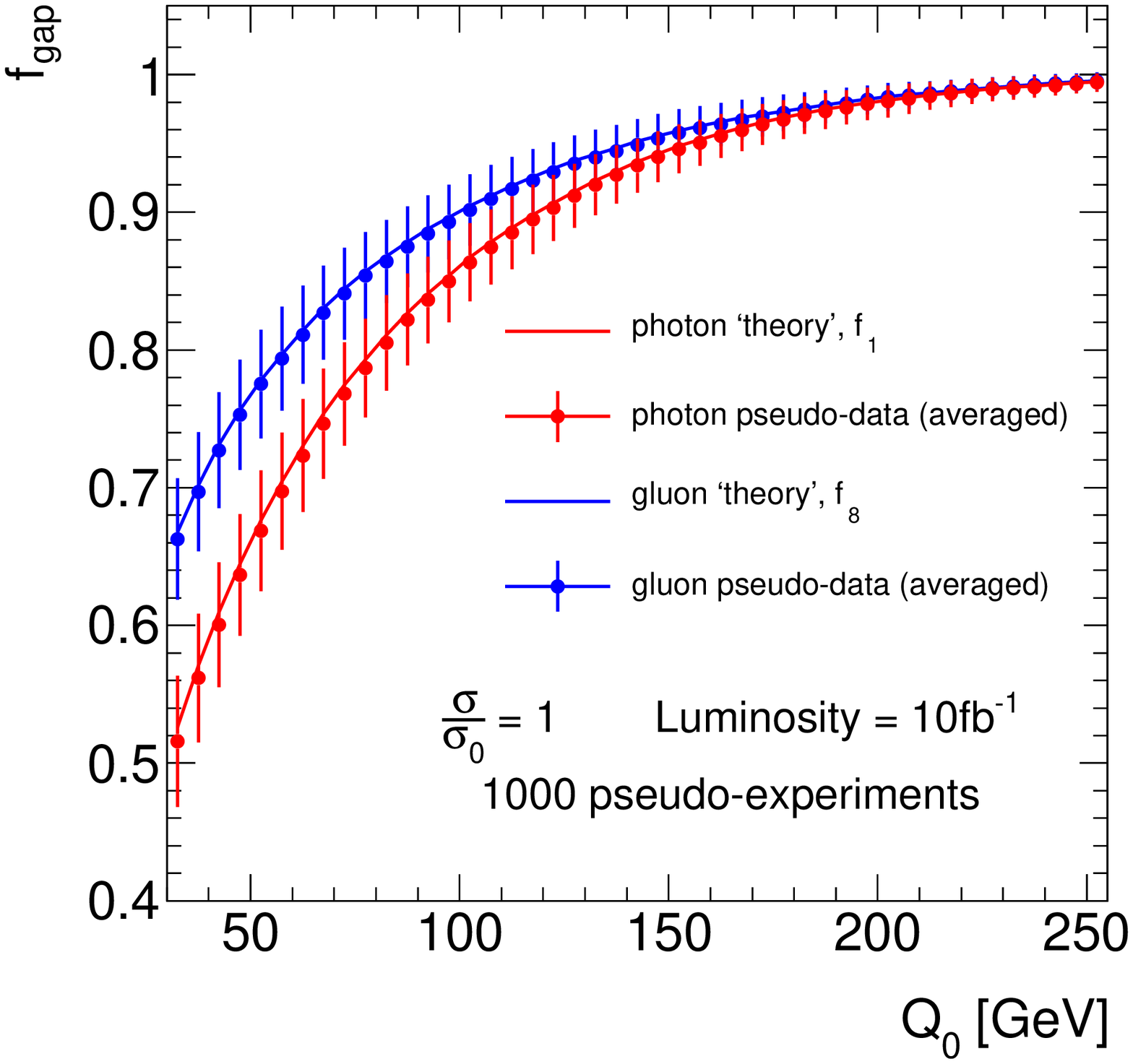} }%
}
\caption{(a) The data points show the gap fraction in a typical pseudo-experiment for
  the case of a heavy gluon. The curves show the corresponding
  theoretical predictions for both a heavy photon and a heavy gluon; 
(b) Mean gap fraction obtained after averaging across 1000 pseudo-experiments. 
The errors in (b) represent the RMS spread of values obtained in the pseudo-experiments. 
\label{fig:gap-fracs}}
\end{figure}

The gap-fraction as a function of $Q_0$ for the signal events can be constructed by
\begin{equation} 
f_{gap} (Q_0) = \frac{N_{s}(Q_0)}{N_s(Q_{0}=300~\mbox{GeV})}.
\label{eq:fgap}
\end{equation}
Figure \ref{fig:gap-fracs}(a) shows the result of a typical
pseudo-experiment in the case of a heavy 
gluon signal and assuming an integrated luminosity of 10~fb$^{-1}$. The pseudo-data are compared to 
the theoretical predictions (solid lines, \pythia8) for the heavy gluon and
heavy photon cases. 
Although there are sizable statistical fluctuations in the 
pseudo-data\footnote{The errors shown in Figure~\ref{fig:gap-fracs}(a) correspond to 
treating the numerator and denominator in Eq.~(\ref{eq:fgap}) as uncorrelated, Poisson, random 
values.}, it is clear that they are better represented by the heavy gluon prediction.

Our ability to extract the signal correctly is confirmed in Figure
\ref{fig:gap-fracs}(b). Here the gap fraction is obtained by computing the
mean across all pseudo-experiments. 
We clearly see that the mean gap fractions agree well with the signal-only
theoretical predictions for both heavy gluon and heavy photon cases. The error bars represent the RMS spread 
of $f_{gap}$ values and indicate that the gap fraction could have significant 
discriminating power with as little as 10~fb$^{-1}$ of data provided
the production rates are not much smaller than we are assuming. 

\subsection{Extracting the colour}    
    
\begin{figure}[h!t]%
\centering

\mbox{
\subfigure[]{\includegraphics[width=0.45\textwidth]{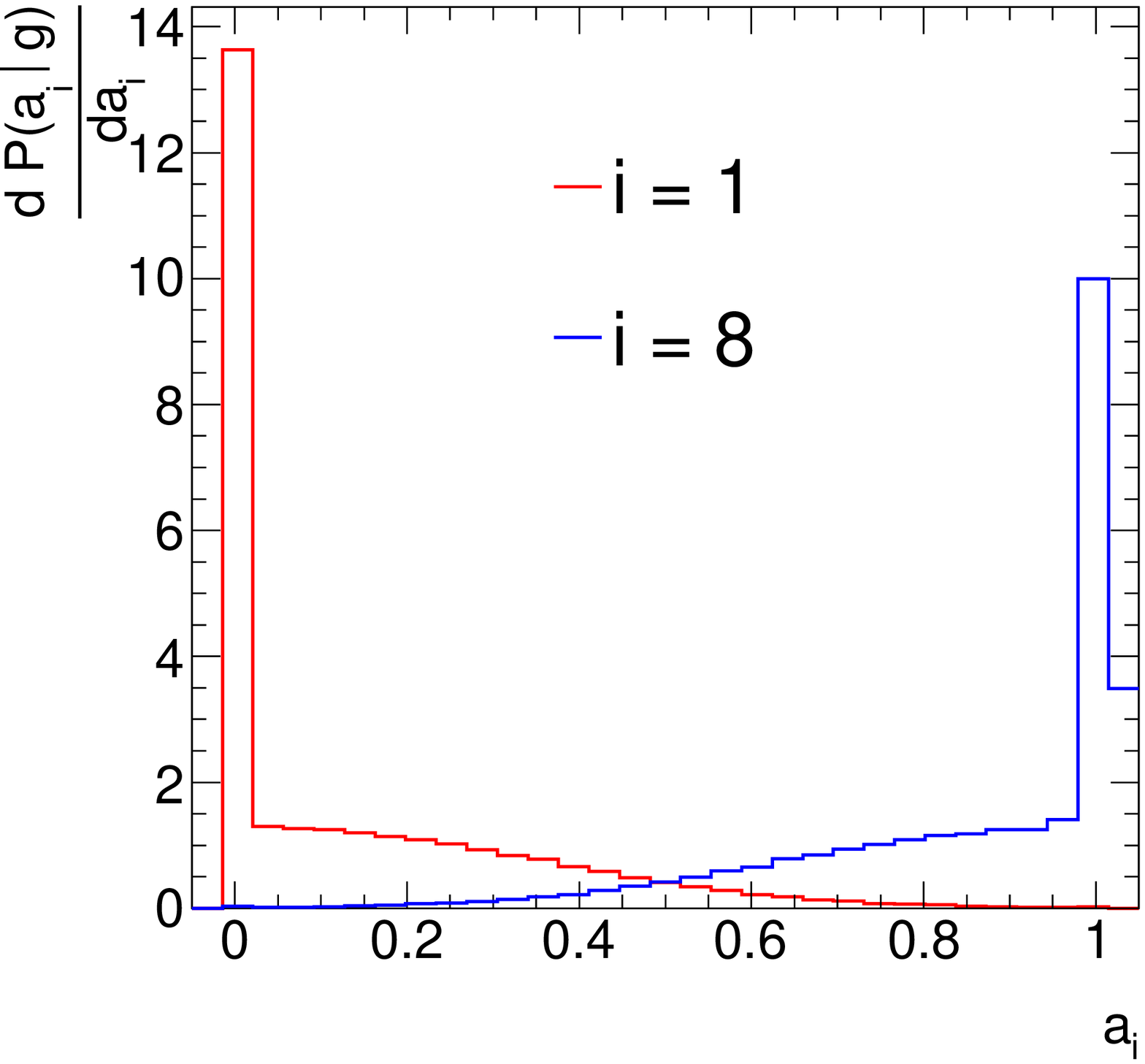} }%
\subfigure[]{\includegraphics[width=0.45\textwidth]{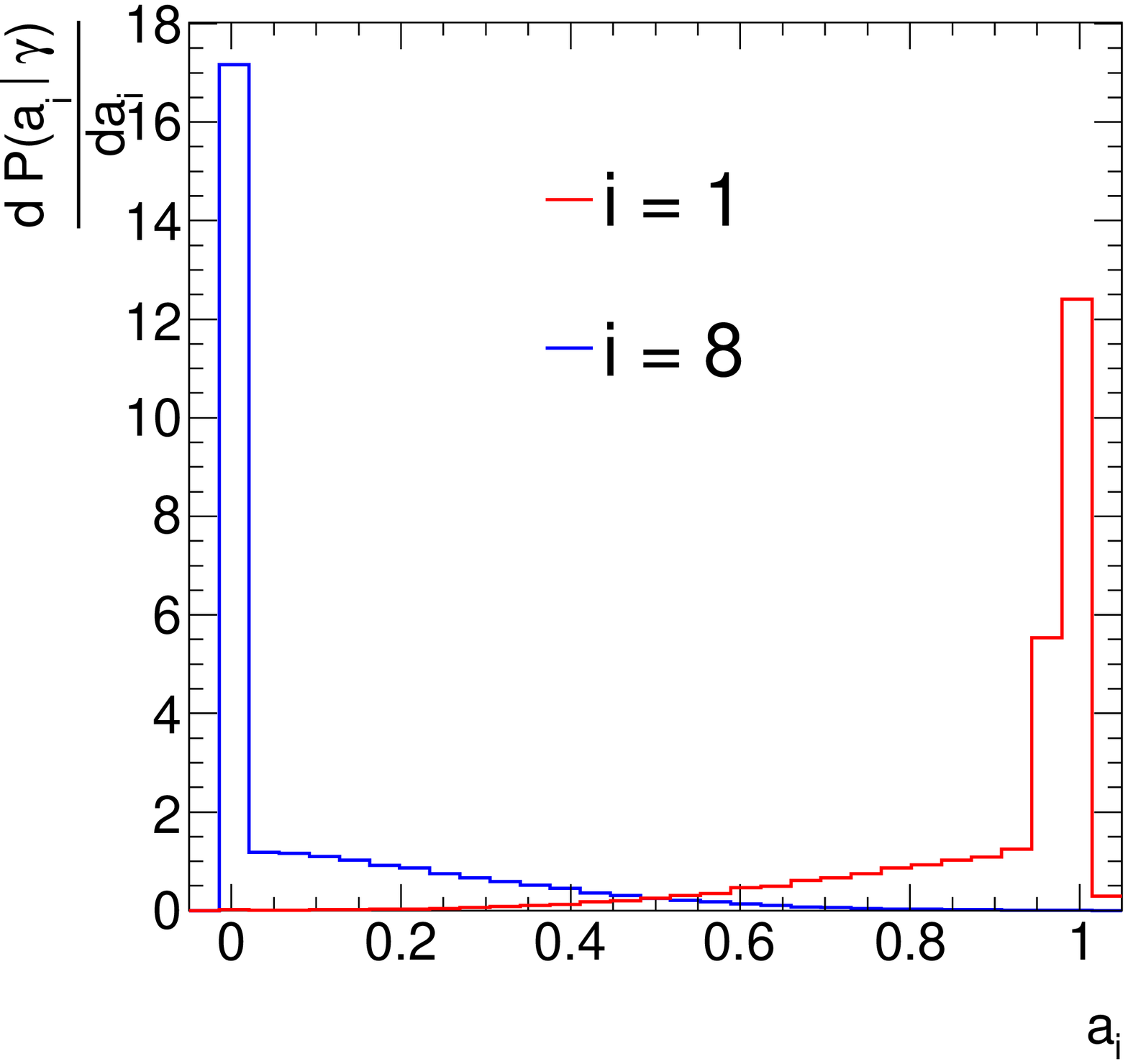} }%
}
\caption{(a) $P(a_1 | \rm g)$ and $P(a_8 | \rm g)$ obtained from
  $10^5$ pseudo-experiments for the case of a heavy gluon resonance. (b) $P(a_1 | \rm \gamma)$ 
and $P(a_8 | \rm \gamma)$ obtained from $10^5$ 
pseudo-experiments for the case of a heavy photon resonance. Assuming
10 fb$^{-1}$ of data.\label{fig:ABdist}}
\end{figure}

The colour of the resonance can be obtained using a fit of the form
\begin{equation}\label{eq:fgapfit}
f_{gap} (Q_0) = a_1 \, f_{1}(Q_0) + a_8 \, f_{8}(Q_0)
\end{equation}
where $f_1$ ($f_8$) is the pure theory prediction for a colour singlet (octet) resonance 
and the $a_i$ are constants that are allowed to vary in the fit to
data\footnote{We constrain them to be positive.}. Figure~\ref{fig:ABdist}(a) shows the probability of obtaining a specific value 
of $a_8$ for a heavy gluon signal, $P(a_8 | \rm g)$, assuming an integrated luminosity 
of 10~fb$^{-1}$. $P(a_8 | \rm g)$ is strongly peaked at unity, indicating that the gluon is 
identified as such in the majority of pseudo-experiments. Also shown
is $P(a_1 | \rm g)$ and, as expected, it is strongly peaked at
zero\footnote{We observe that $a_1+a_8 \approx 1.0$ with an RMS
 variation of less than 2\%}. Figure \ref{fig:ABdist}(b) shows the corresponding distributions 
when the true signal is a heavy photon, i.e. $P(a_1 | \rm \gamma)$ and $P(a_8 | \rm \gamma)$.

\begin{figure}[h!t]%
\centering

\mbox{
\subfigure[]{\includegraphics[width=0.45\textwidth]{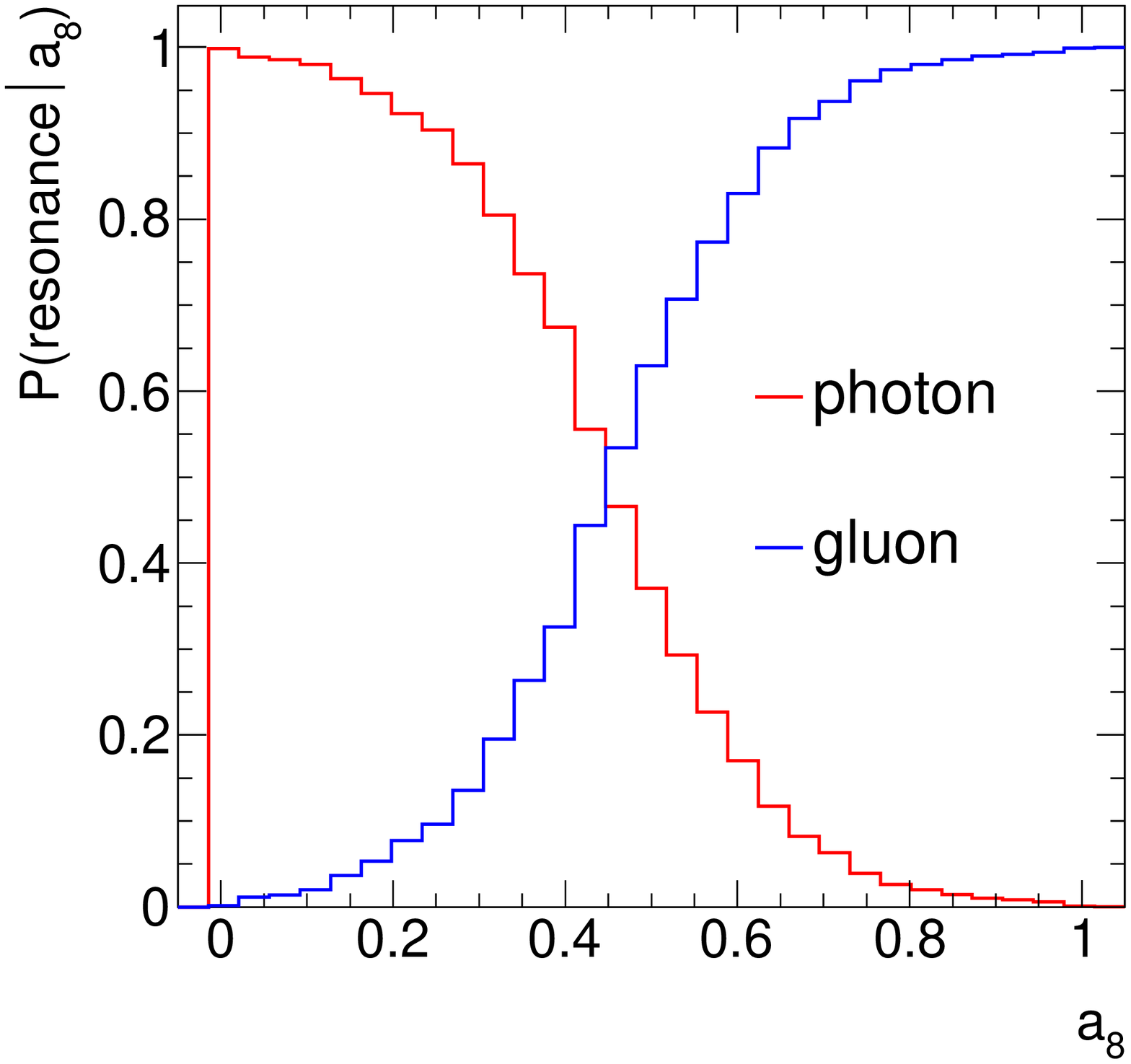} }%
\subfigure[]{\includegraphics[width=0.45\textwidth]{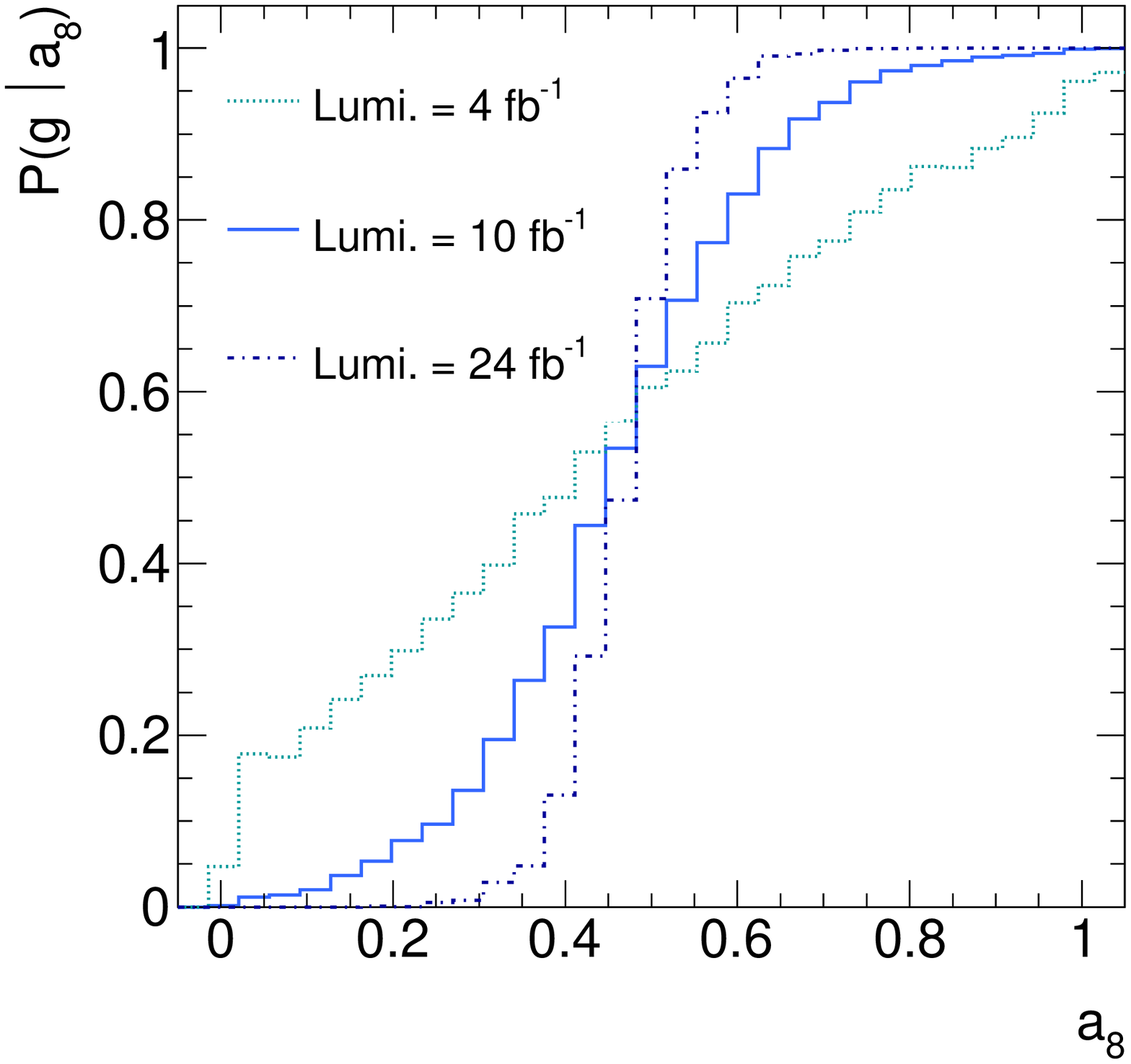} }%
}

\caption{(a) $P(\rm g | a_8)$ and $P(\rm \gamma | a_8)$ in the case of a heavy gluon resonance 
and assuming the baseline signal size and 10~fb$^{-1}$ of integrated luminosity. (b) $P(\rm g | a_8)$ 
for three different values of the luminosity.
\label{fig:probs}}
\end{figure}

More interesting is the probability that a particular measured value
of $a_8$ is due to a gluon resonance and it
can be calculated using Bayes' Theorem if we assume that prior to
this analysis the resonance is equally likely to be a heavy gluon or
a heavy photon:
\begin{equation}
P({\rm g} | a_8) = \frac{P(a_8 | {\rm g})}{P(a_8 | {\rm g}) + P(a_8 | {\rm \gamma})}.
\end{equation}
Figure \ref{fig:probs}(a) shows the values of $P({\rm g} | a_8)$ and $P({\rm \gamma} | a_8)$ 
as a function of $a_8$. As expected, $P({\rm g} | a_8)$ increases as $a_8$ increases, 
whereas the $P({\rm \gamma} | a_8)$ decreases. Figure \ref{fig:probs}(b) shows the 
values of $P({\rm g} | a_8)$ for different values of the input luminosity. 

Thus, for each pseudo-experiment a specific value of $a_8$ can be extracted
from the gap fraction fit and the probability 
that the resonance is a gluon (or a photon) can be determined using the probability 
distributions shown in Figure~\ref{fig:probs}: Exactly the same procedure could be 
used with real data. To quantify the feasibility of making such a measurement at the 
LHC, we calculate the fraction of pseudo-experiments that have $P({\rm g} | a_8)$ larger 
than 95\%, which we denote $G_{95}$. For example,  $G_{95}=0.77$ for the baseline heavy 
gluon signal and 10~fb$^{-1}$ of integrated luminosity. It is very likely, therefore, 
that the LHC experiments could identify the colour structure of such a 
resonance, should it occur in Nature.

\begin{figure}[h!t]%
\centering

\mbox{
\subfigure[]{\includegraphics[width=0.45\textwidth]{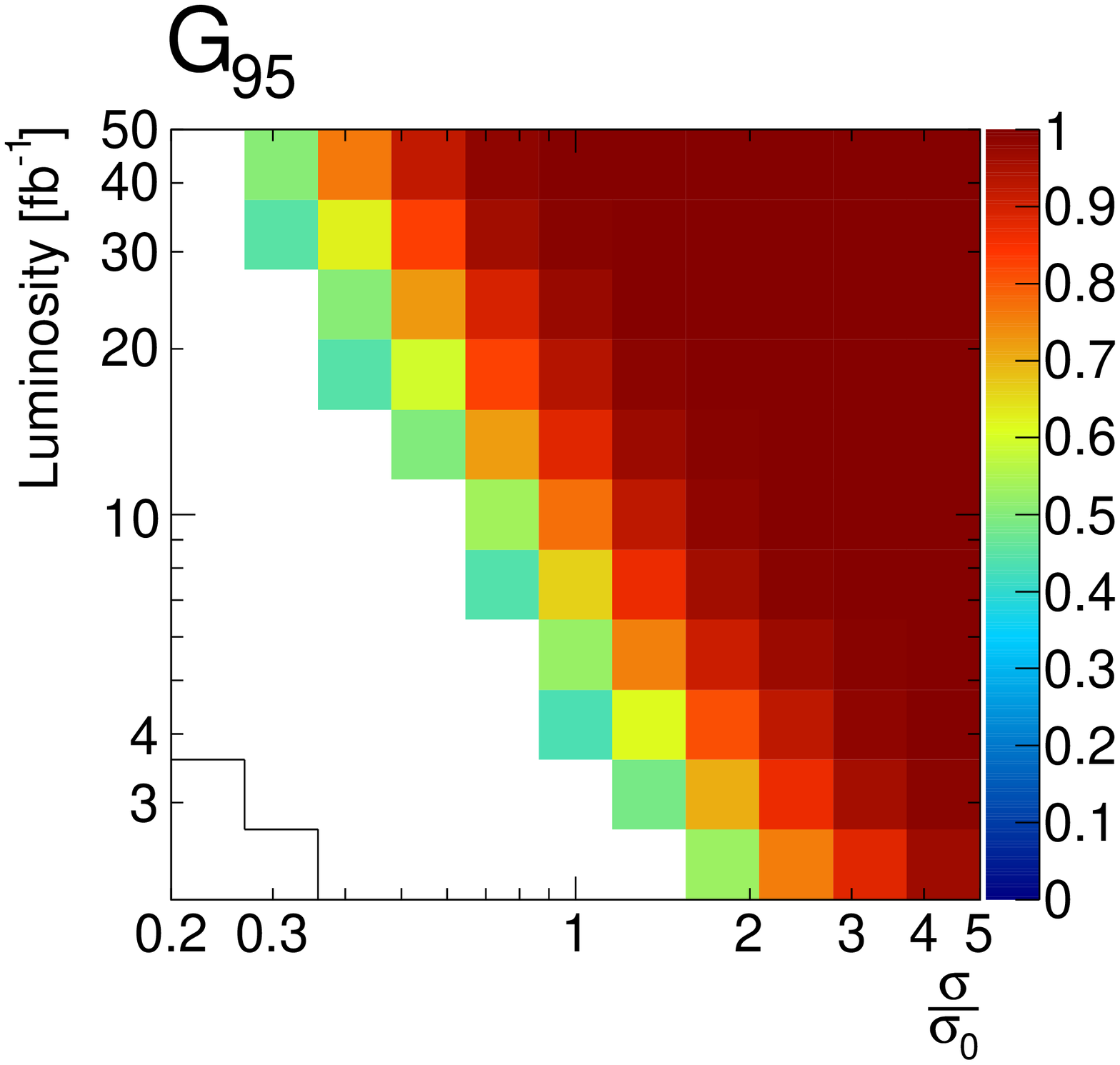}}\quad
\subfigure[]{\includegraphics[width=0.45\textwidth]{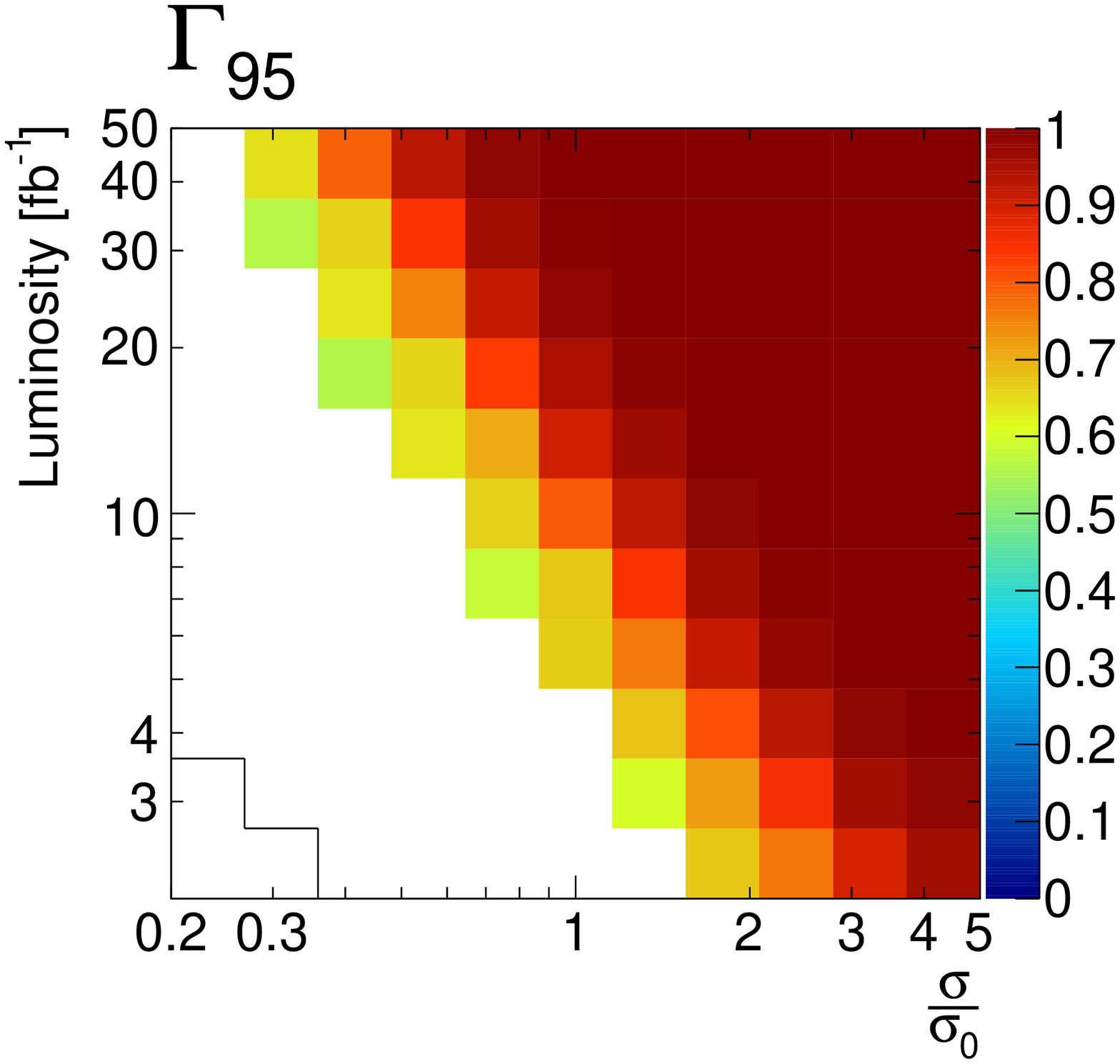}}%
}

\caption{The fraction of LHC experiments that would measure $P({\rm g} | a_8)\geq 95\%$ 
is denoted by $G_{95}$: (a) $G_{95}$ as a function of the luminosity and 
signal size assuming a heavy gluon resonance. The fraction of LHC experiments that 
would measure $P({\rm \gamma} | a_1)\geq 95\%$ is denoted by
$\Gamma_{95}$: (b) $\Gamma_{95}$ as a function of the luminosity and
signal size assuming a heavy photon resonance. The solid line in the bottom left corner indicates the region where 
the significance of any signal is less than $5\sigma$ significance.
\label{fig:LvsS_gluon}}
\end{figure}

\begin{figure}[h!t]%
\centering

\mbox{
\subfigure[]{\includegraphics[width=0.45\textwidth]{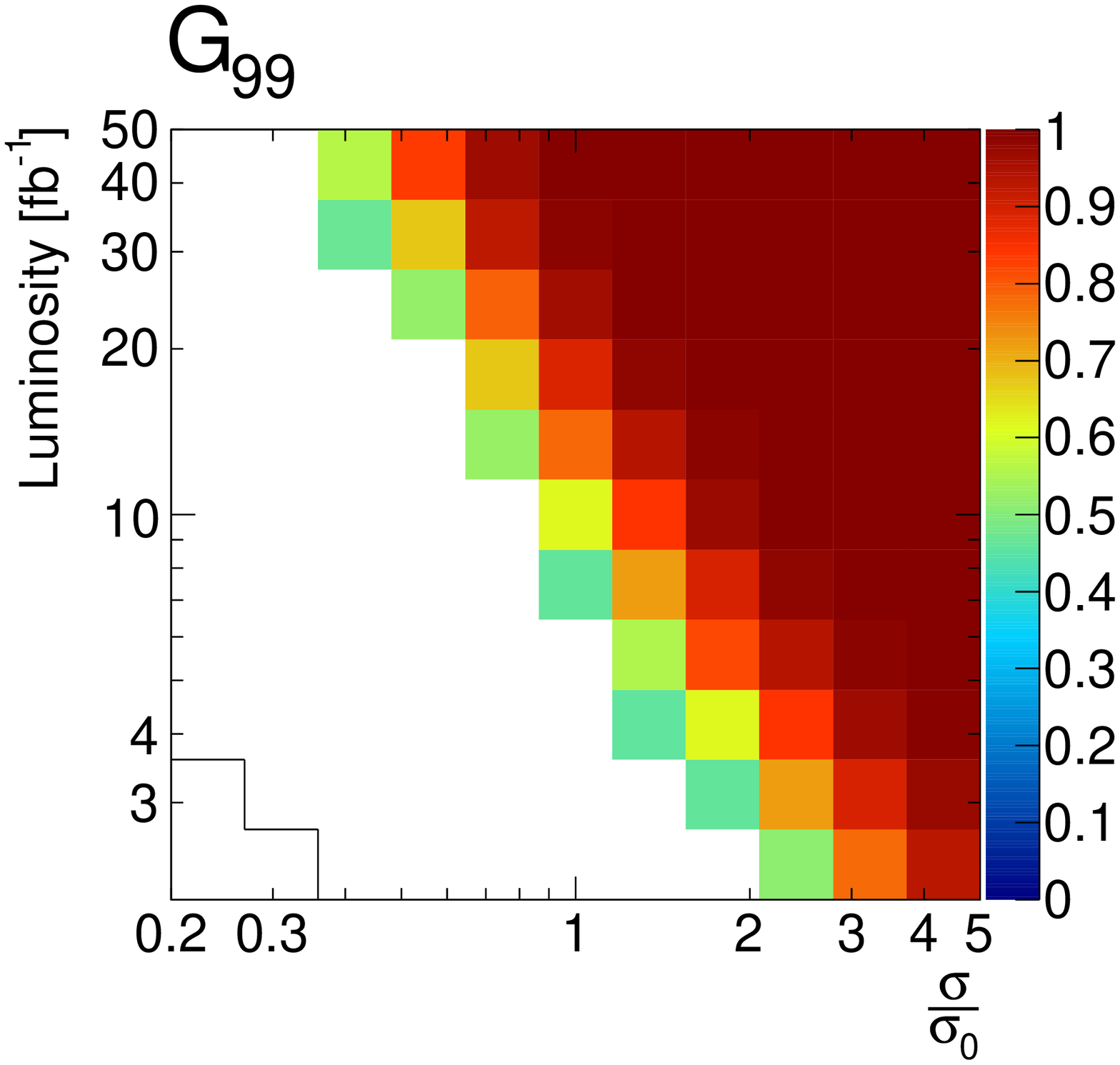}}\quad
\subfigure[]{\includegraphics[width=0.45\textwidth]{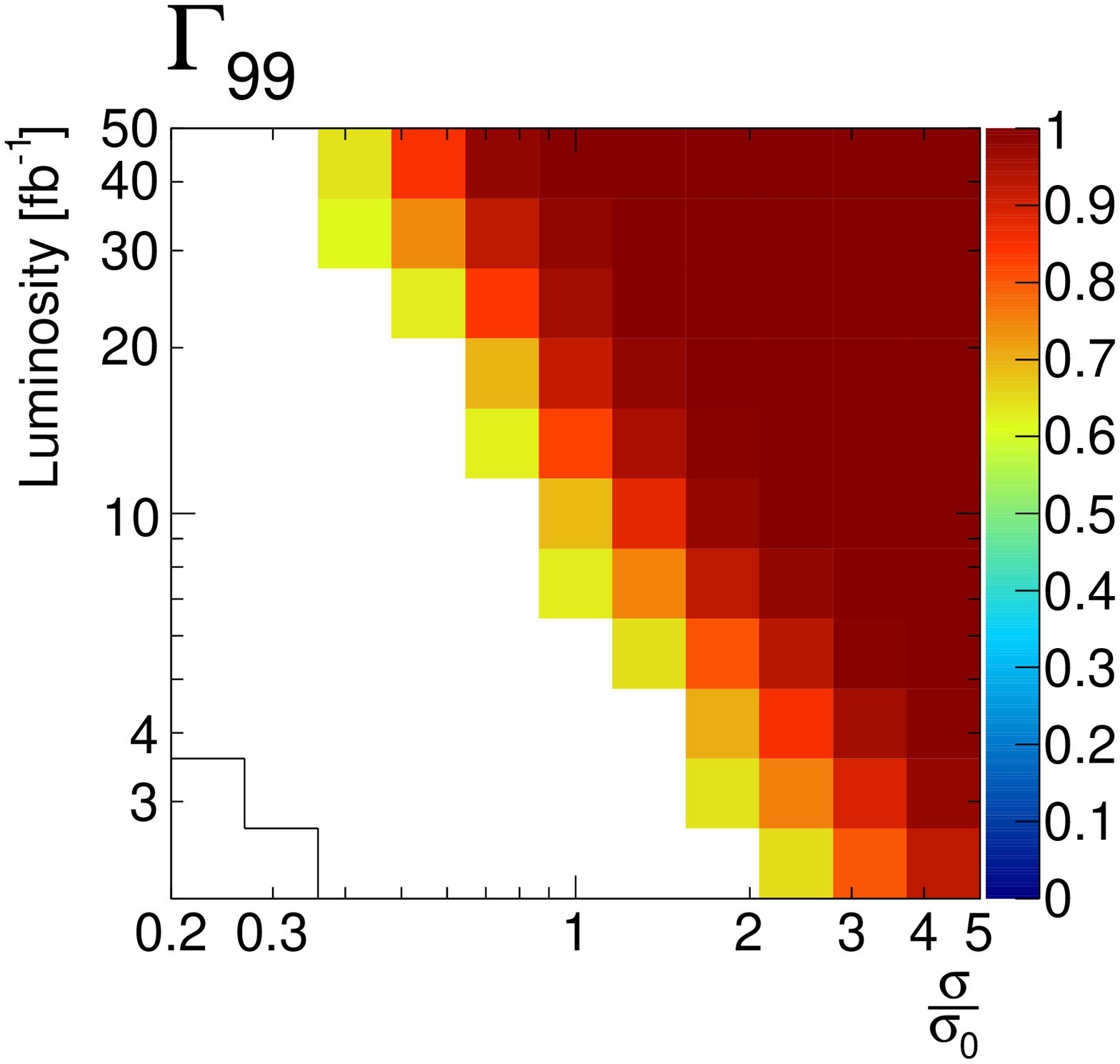}}%
}

\caption{The fraction of LHC experiments that would measure $P({\rm g} | a_8)\geq 99\%$ 
is denoted by $G_{99}$: (a) $G_{99}$ as a function of the luminosity and 
signal size assuming a heavy gluon resonance. The fraction of LHC experiments that 
would measure $P({\rm \gamma} | a_1)\geq 99\%$ is denoted by
$\Gamma_{99}$: (b) $\Gamma_{99}$ as a function of the luminosity and
signal size assuming a heavy photon resonance. The solid line in the bottom left corner indicates the region where 
the significance of any signal is less than $5\sigma$ significance.
\label{fig:LvsS_gluon2}}
\end{figure}

In Figure \ref{fig:LvsS_gluon}(a) we show how
$G_{95}$ varies as we vary the size of the signal cross section
relative to the baseline value ($\sigma_0$) and for different
integrated luminosities. For signal cross sections less than around $1/3$ of the
baseline value, a luminosity in excess of 50~fb$^{-1}$ will be needed in order to extract the colour 
the resonance. Figure \ref{fig:LvsS_gluon}(b) shows the equivalent
distribution but now for a heavy photon resonance
($\Gamma_{95}$). Figure \ref{fig:LvsS_gluon2} shows the corresponding
plots when the probability per pseudo-experiment is increased to 99\%
($G_{99}$ and  $\Gamma_{99}$) and it is encouraging to note that the
plots look rather similar to the previous case. 

\subsection{Effect of experimental and theoretical uncertainties \label{sec:syst}}

Experimental and theoretical uncertainties will obviously affect 
our ability to determine the colour of the resonance. The experimental uncertainties 
associated with the extraction of the signal will, to a large extent, cancel in the 
ratio and so will not affect the measurement in a significant way. These include the 
top-tagging efficiency, luminosity and energy scale/resolution of the top-tagged jets. 
However, the uncertainties associated with the veto jets, i.e. the jet energy scale, jet energy 
resolution and jet reconstruction efficiency, do affect our ability to make the measurement. 
A feel for the likely systematic uncertainty from the experiment can be obtained
using the ATLAS 
measurement of dijet production with a veto on additional central jet activity \cite{bib:atl11}. 
In particular, the total systematic uncertainty on the $f_{gap}(Q_0)$ distributions were 
at worst 5\% for $Q_0=20$~GeV for dijet systems constructed from central high-$p_{\rm T}$ 
jets.

The current theoretical uncertainties are probably much
larger than this. For example, the theory 
predictions in \cite{bib:atl11} deviate from the ATLAS dijet data by
$\sim 25\%$ at $Q_0=20$~GeV and in \cite{bib:del11} the uncertainty in
the theoretical prediction of the gap fraction was found to be
approaching 50\% in some regions of phase space. There is no reason to
suppose that jet vetoing in boosted $t\bar{t}$ production is much
better understood. Indeed, the difference between Madgraph and
\pythia8 was shown in Section \ref{sec:veto} to be about 10\%. It is clear, therefore, that the theoretical uncertainties are the limiting factor in the 
feasibility of making this measurement.

\begin{figure}[h!t]%
\centering
\mbox{
\subfigure[]{\includegraphics[width=0.45\textwidth]{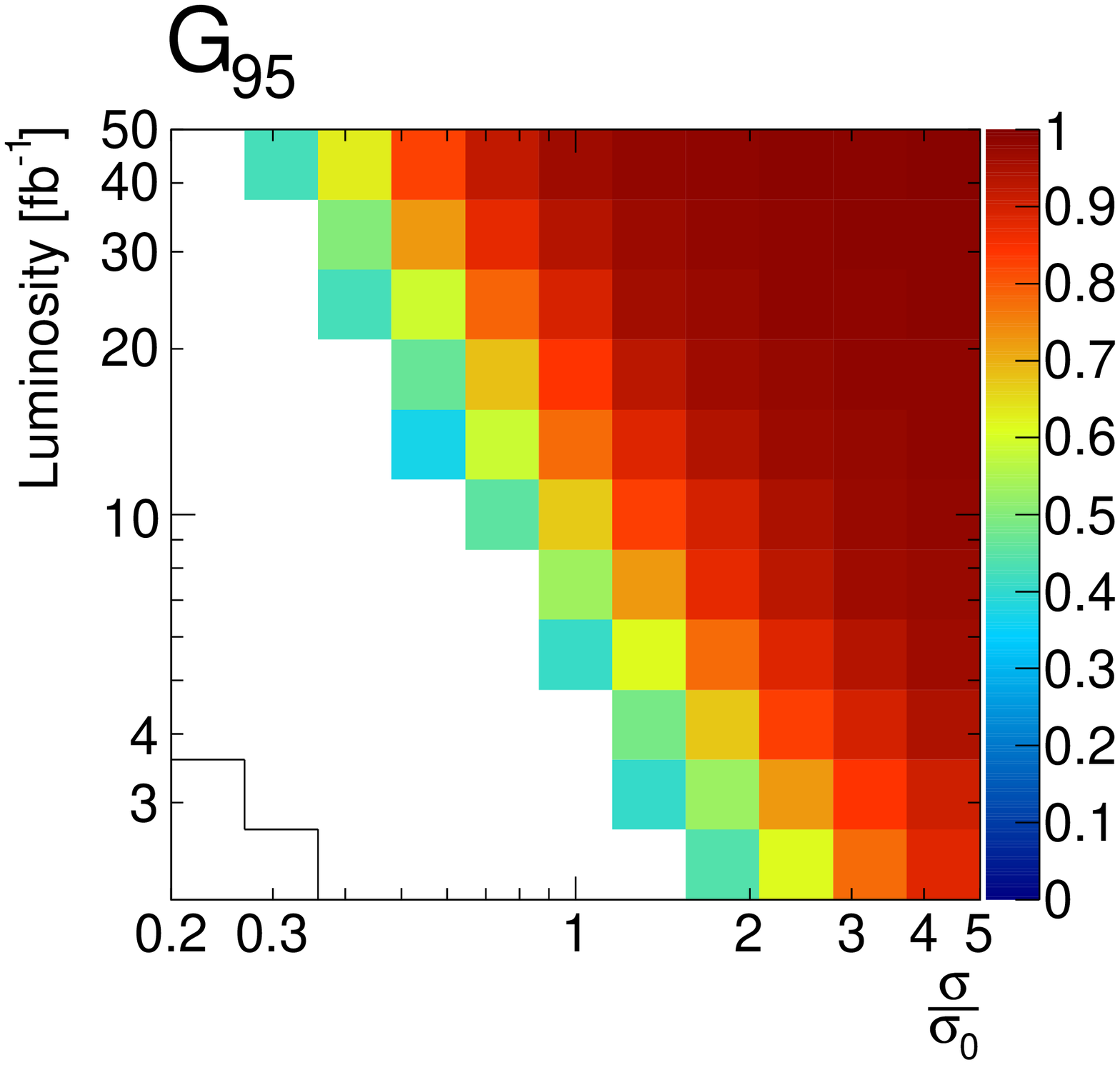}}\quad
\subfigure[]{\includegraphics[width=0.45\textwidth]{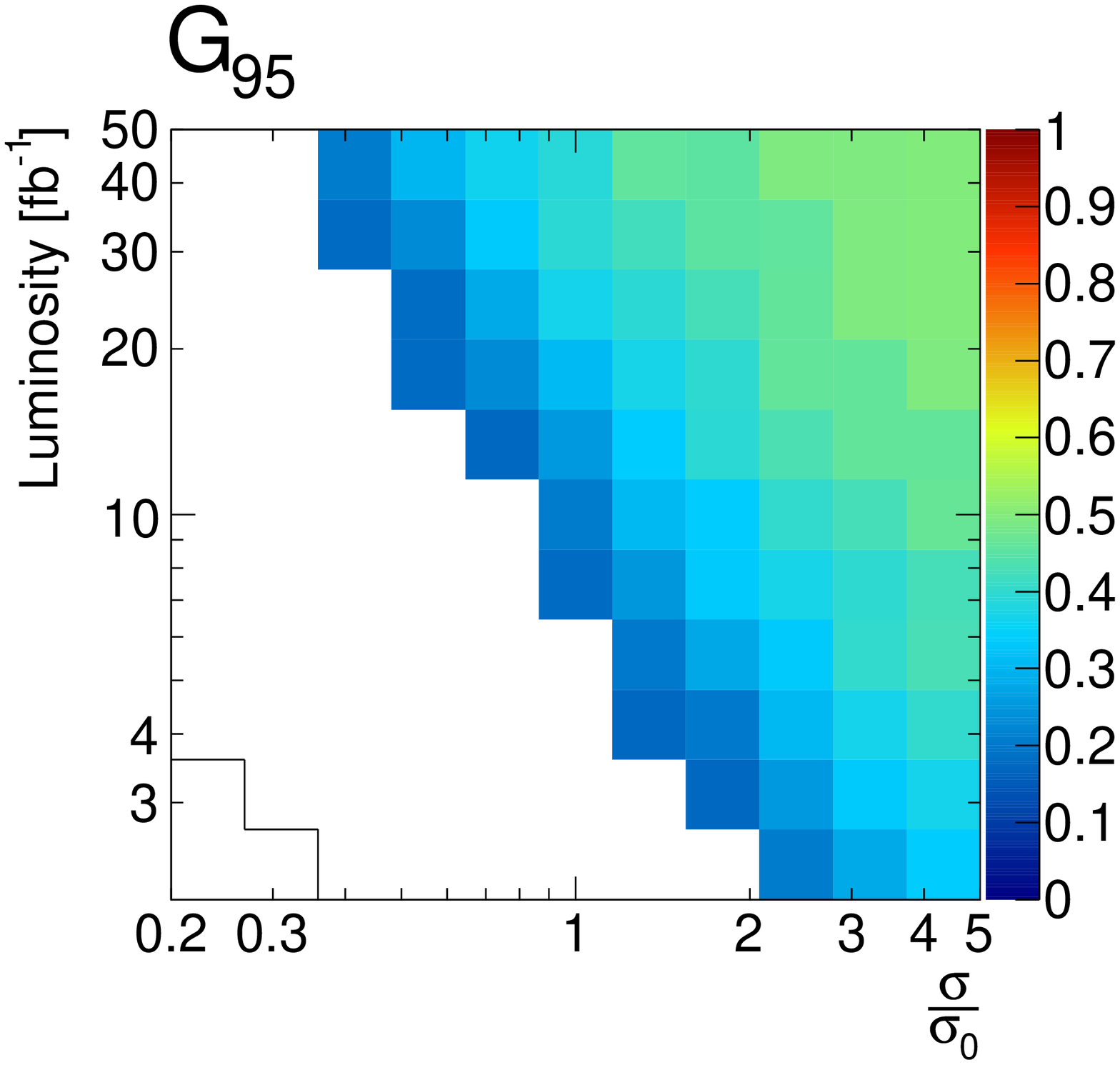}}%
}
\caption{(a) The impact of a 10\% uncertainty in the gap fraction shape on the value of 
$G_{95}$. (b) The impact of a 25\% uncertainty in the gap fraction shape on the value of 
$G_{95}$. \label{fig:syst1}}
\end{figure}

We quantify the impact of the theory uncertainty by arbitrarily changing the shape of the theoretical 
predictions used to fit the pseudo-data (i.e. $f_1$ and $f_8$). In each pseudo-experiment, 
we choose two uniformly distributed random numbers in the interval $[-X, +X]$, which 
give the fractional shifts that are applied to $f_1$ and $f_8$ at $Q_0=20$~GeV. The shift of each gap fraction  
at $Q_0=300$~GeV is by definition zero and for $Q_0$ values between
20~GeV and 300~GeV the shift is obtained 
by a linear interpolation. Figure \ref{fig:syst1} shows $G_{95}$ for
$X = 10$\% and 25\% and an uncertainty in the gap fraction 
at the 25\% level clearly has a major impact on the measurement. Encouragingly, an overall 10\% 
uncertainty does not degrade the measurement much. It is therefore clear that the 
theoretical uncertainty should be reduced to around the 10\% level 
in order to be confident of using  jet vetoing as a tool to extract the colour of heavy
TeV-scale resonances.

\section{Acknowledgments}
We would like to thank Phil Bull and Christina Smith, for
their insights on top-tagging, and Johan Alwall for help with Madgraph.
This work was funded in the UK by STFC and in part by the EU Marie Curie Research Training 
Network ``MCnet'', under contract number MRTN-CT-2006-035606. JC would like to acknowledge 
the support received for this work from the Muir Wood Award Fund granted by Peterhouse in 
Cambridge.

\bibliographystyle{JHEP}
\bibliography{resonance}

\appendix 
\section{Appendix\label{app:proc}}

This appendix highlights some of the features 
of the heavy gluon implementation in \pythia8. In particular we take a
look at the strong enhancement of the $t\bar{t}$ invariant mass distribution
at low masses due to parton distribution function (PDF) effects,
the potentially large interference effects that can occur between the
signal and background and the forward-backward
asymmetry that can be induced by non-chirally symmetry couplings of
the heavy gluon to quarks. 
 
For this appendix only, we take as our reference 
the typical RS couplings: $g^q_L = g^q_R = g^b_R = -0.2$, $g^b_L = g^t_L = 1$, $g^t_R = 4$ 
\cite{bib:lil07}. These contain a significant axial-vector coupling 
to tops\footnote{We define $g_v = (g_L + g_R)/2$ and $g_a = (g_L - g_R)/2$.}. 
In the following, we do also vary the couplings about these reference values. 
Also for this appendix only, when we speak of the $t\bar{t}$ mass we refer to the 
invariant mass of the top pair at parton level, \ie\ before showering and hadronisation.

\begin{figure}[h!t]
\centering
\mbox{
 \subfigure[]{\includegraphics[width=0.45\textwidth]{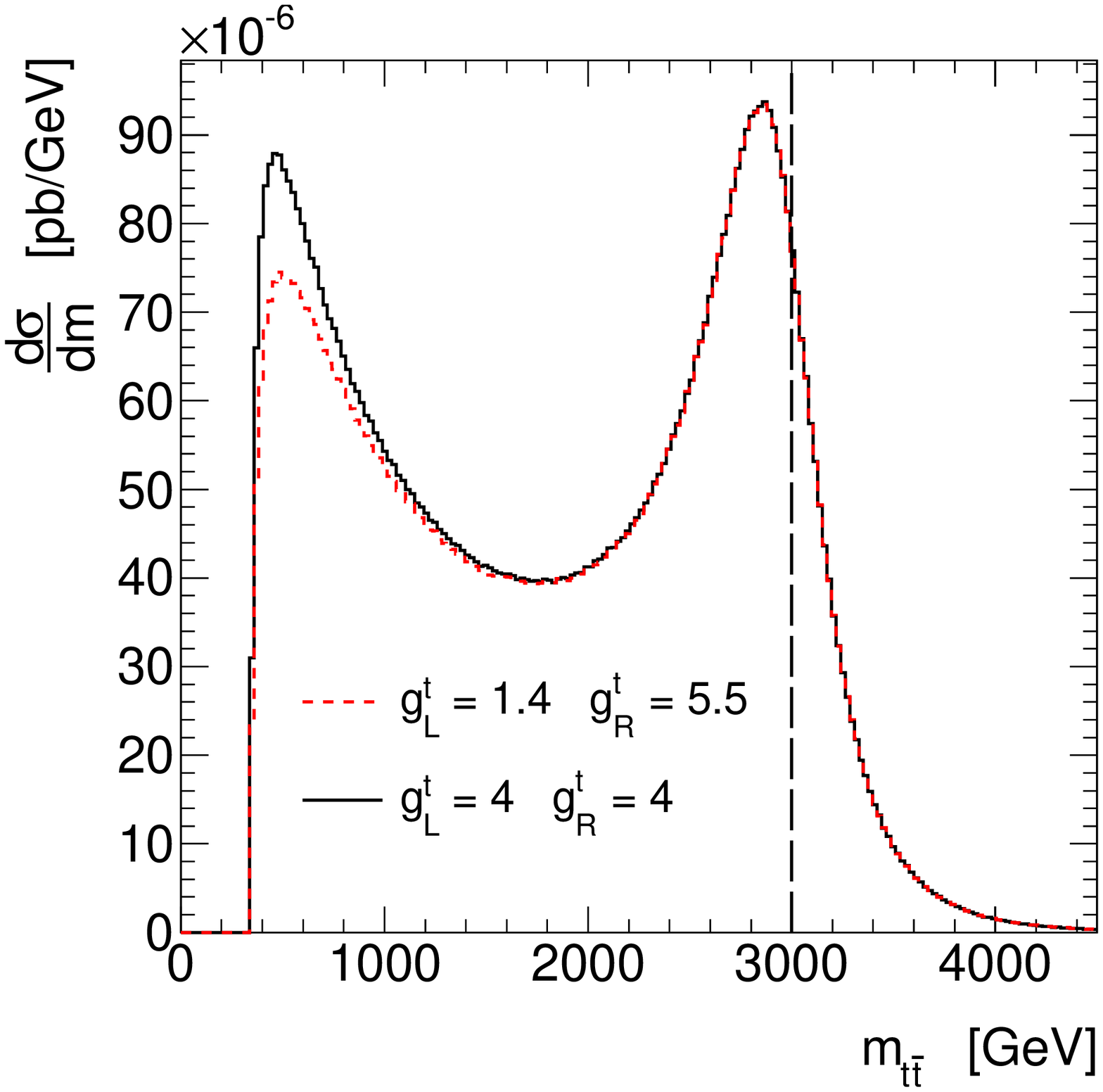}}
 \subfigure[]{\includegraphics[width=0.45\textwidth]{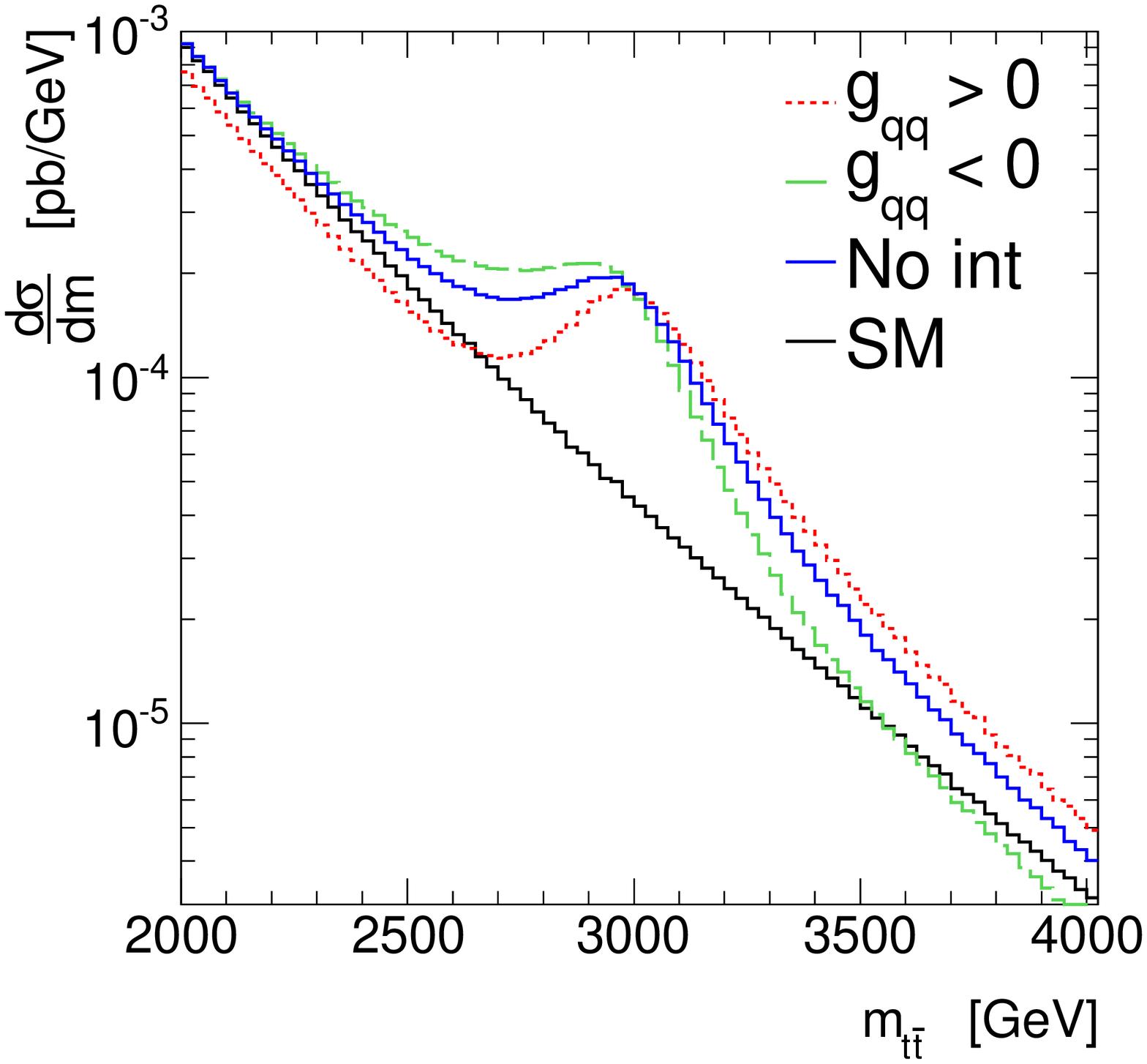}}
}
\caption{(a) Invariant $t\bar{t}$ mass distribution for a broad gluon 
resonance, neglecting the SM contribution. (b) Comparison of $t\bar{t}$  invariant 
mass distributions for several RS-like scenarios. Contributions from
heavy and SM gluon production are included, both from $q\bar{q}$ annihilation and gluon fusion processes. 
The details are explained in the text.\label{fig:app1}}       
\end{figure}

The first point of note is that the shape of a wide resonance can be very strongly 
distorted by the PDFs of the incoming particles. 
The fast increase of the PDFs with decreasing $x$ competes with the suppression from the 
heavy gluon virtuality, enhancing the low mass tail of the resonance and suppressing it 
at high masses. This is illustrated in Figure \ref{fig:app1}(a), which shows the 
$t\bar{t}$ mass spectrum for the `pure' heavy gluon process (\ie\ without any SM background  
or interference effects) for a wide resonance with $m_{\gkk} = 3$~TeV, $g^t_v = 4$ and 
$g^t_a = 0$. The shift of the peak can be significant for large resonance mass and 
couplings, but tends not to be larger than about 100~GeV for typical RS scenarios that 
might be seen at the LHC. Similarly, large mass and couplings also imply that a large 
fraction of the resonance cross section resides in the low mass tail and this fraction 
is sensitive to the choice of mass and couplings. It should also be noted that the tail 
fraction can be significant even for a relatively light resonance and that it is quite 
possible to encounter scenarios where only the minority of the cross section resides 
in the resonance peak. Figure~\ref{fig:app1}(a) also demonstrates the small difference 
introduced by adding some axial coupling while maintaining the same overall normalisation 
($g^t_v = 3.45$, $g^t_a = 2.05$). It can be seen that while the balance of axial and 
vector couplings does have some effect on the low mass tail of the $t\bar{t}$ mass 
spectrum, its effect on the main part of the resonance peak is very small.

\begin{figure}[h!t]
\centering
\mbox{
  \subfigure[]{\includegraphics[width=0.45\textwidth]{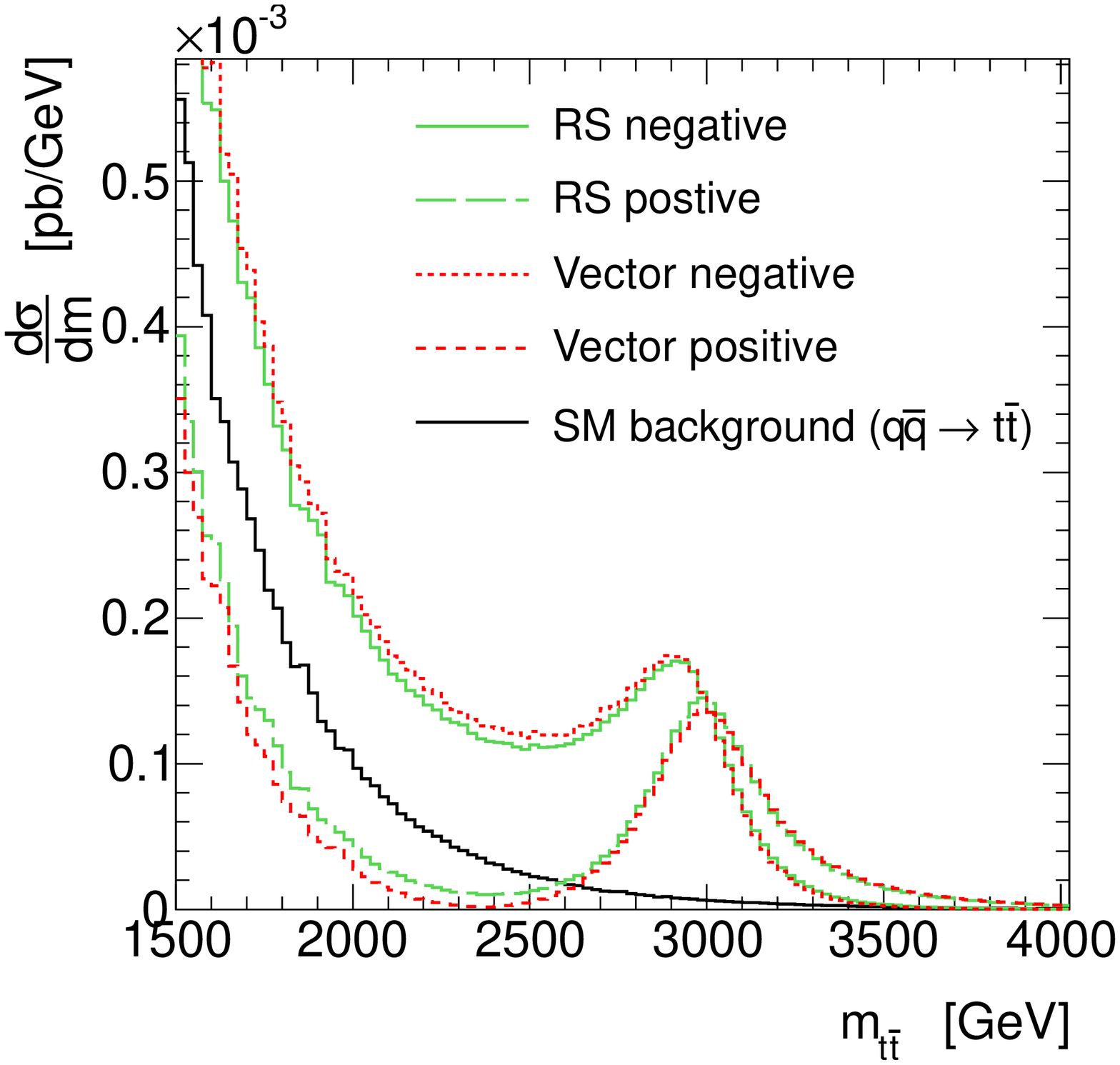}}
  \subfigure[]{\includegraphics[width=0.45\textwidth]{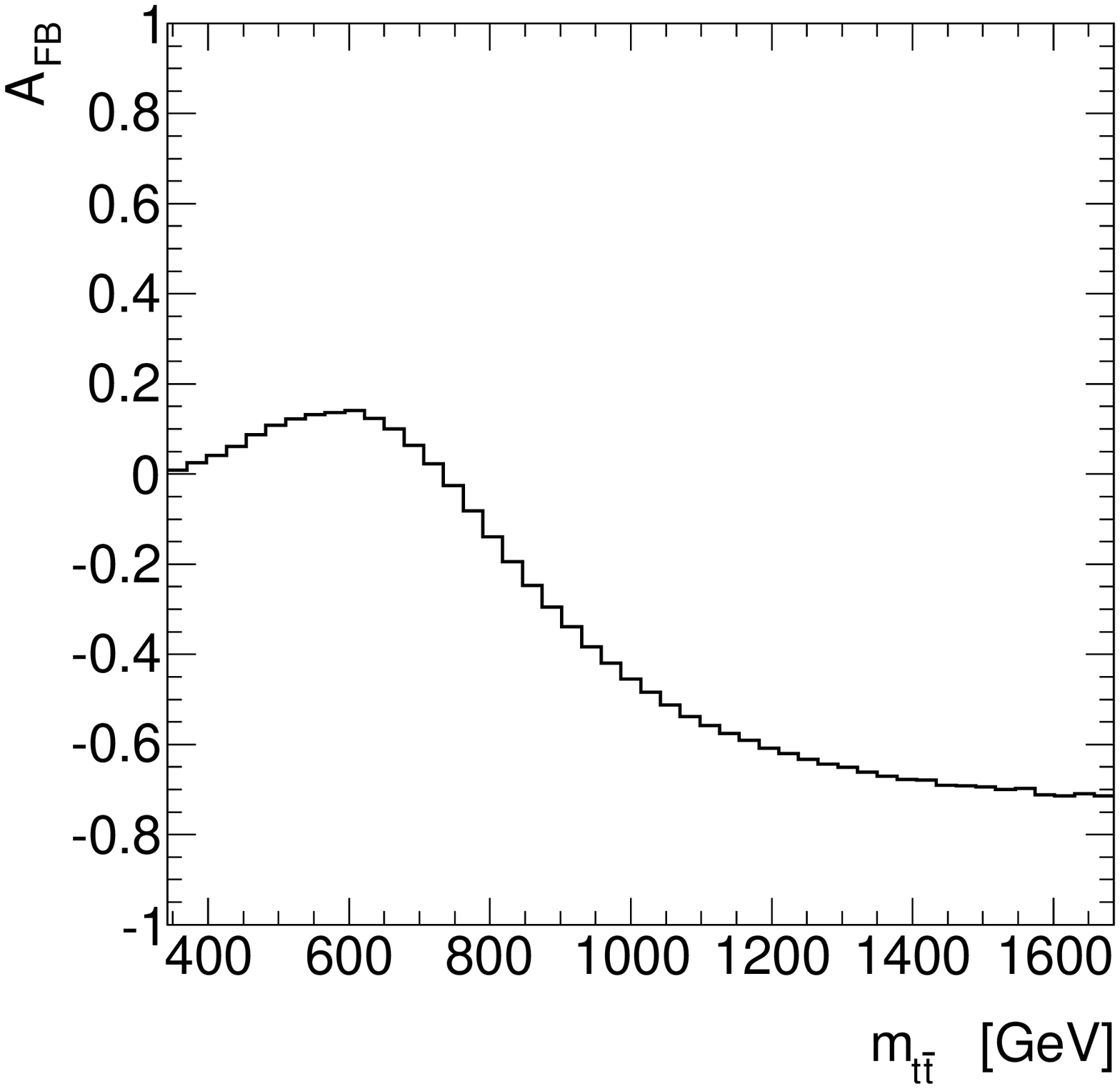}}
}
\caption{(a) Comparison of invariant $t\bar{t}$ mass distributions for RS-like and 
vector coupling scenarios. Contributions from both heavy and SM gluon production are included, 
but only from $q\bar{q}$ annihilation. (b) Forward-backward asymmetry as a function of 
the $t\bar{t}$ invariant mass. The details are explained in the text.\label{fig:app2}}       
\end{figure}

While most of the SM $t\bar{t}$ events at the LHC are initiated by gluon fusion, 
a significant fraction comes from quark annihilation, which can interfere with the 
heavy gluon amplitude. The nature of this interference is sensitive to the model 
parameters and so, in principle, could be used to constrain the
heavy gluon. Note that the interference changes sign at the \gkk\ mass: RS models 
typically predict a negative coupling to the incoming light quarks and a positive 
coupling to the outgoing top quarks, which would result in constructive interference 
below the resonance peak and destructive interference above. This would be reversed 
if the couplings to the incoming and outgoing quarks were of the same sign. 
Figure~\ref{fig:app1}(b) compares the $t\bar{t}$ mass spectrum for the typical RS 
scenario ($g_{qq}<0$) to an equivalent scenario with only positive couplings 
($g_{qq}>0$) and also to the situation where interference effects are not taken 
into account (``No int''). As pointed out, for example in \cite{bib:lil072}, this shows 
that interference effects can certainly be significant, even on top of the full SM 
$t\bar{t}$ background. We can say a little more about how the strength of the 
interference depends on the balance of axial and vector top couplings: Since only the 
vector part of the heavy gluon coupling interferes with the SM background it is to be 
expected that reducing the vector coupling and increasing axial coupling will reduce 
the strength of the interference. Figure~\ref{fig:app2}(a), which compares the RS 
scenario with interference that was used to produce Figure \ref{fig:app1}(b) ($g^{t}_{v} = 2.5, g^{t}_{a} = 1.5$) 
to its purely vector coupling equivalent ($g^{t}_{v} = 2.92, g^{t}_{a} = 0$), 
demonstrates that this difference is small when comparing RS-like to vector scenarios 
and it will be quite negligible once the gluon fusion background is added in (it is 
not included in the figure for clarity). 

To conclude this appendix, we take a look at the forward-backward asymmetry for $p\bar{p}$ 
collisions at $\sqrt{s} = 1.96$ TeV. Here we can make direct
comparisons with results in the literature in order to validate our 
implementation. Two different sets of parameters were considered:
\begin{itemize}
\item{$m_{\gkk} = 2.75$ TeV, 
$g^q_L = 4.5$, 
$g^q_R = -0.2$, 
$g^b_R = 2$, 
$g^b_L = g^t_L = 0$ 
and $g^t_R = 6.1$ 
based on \cite{bib:djo09};}
\item{$m_{\gkk} = 1.0$ TeV, 
$g^q_v = g^b_v = g^t_v = 0$ 
and $g^q_a = -g^b_a = -g^t_a = 1$ 
based on \cite{bib:fer09}.}
\end{itemize}
For the heavy gluon contribution to the asymmetry, we find $A_{FB} =
0.05$ and $A_{FB} = 0.14$ respectively (at parton level), which
agree with the published results. Figure~\ref{fig:app2}(b) 
shows $A_{FB}$ as a function of the invariant mass of the top pair, using the 
parameter set from \cite{bib:djo09}. We note that
the asymmetry correctly passes through zero at $m_{t\bar{t}} = m_{\gkk}/\sqrt{1+2g^q_vg^t_v}$.

\end{document}